\pdfoutput=1

\documentclass[twocolumn,prd,reprint,preprintnumbers,amsmath,amssymb,superscriptaddress,nofootinbib]{revtex4}

\usepackage{epsfig,psfrag,graphics,verbatim}
\usepackage{graphicx}
\usepackage{dcolumn}
\usepackage{bm}
\usepackage{slashed}
\usepackage{color}
\usepackage{amsmath}
\usepackage[]{units}
\usepackage{ulem}

\usepackage{aas_macros}
\usepackage{url}
\usepackage{hyperref}

\newcommand{\D}{\text{d}}
\newcommand{\dpc}{\delta_{\psi\chi}}
\newcommand{\dnp}{\delta_{\nu\psi}}
\newcommand{\dNc}{\delta_{N\chi}}
\newcommand{\dSc}{\delta_{S\chi}}

\begin{document}

$\,$
\vfill
\break
\vspace*{-45mm}
\begin{flushright}
TUM-HEP 1041/16\\ \vspace{-0.05cm}
CP3-Origins-2016-017 DNRF90\\ \vspace{-0.05cm}
\end{flushright}
\vspace*{-0.6cm}

\title{Gamma-ray triangles:\\ a possible signature of asymmetric dark matter in indirect searches}

\author{Alejandro Ibarra}
\affiliation{Physik-Department T30d, Technische Universit\"at M\"unchen, James-Franck-Stra\ss{}e, D-85748 Garching, Germany}

\author{Sergio Lopez-Gehler}
\affiliation{Physik-Department T30d, Technische Universit\"at M\"unchen, James-Franck-Stra\ss{}e, D-85748 Garching, Germany}
\affiliation{Excellence Cluster Universe, Technische Universit\"at M\"unchen, Boltzmannstra\ss{}e 2, D-85748, Garching, Germany}

\author{Emiliano Molinaro}
\affiliation{CP$^{3}$-Origins, University of Southern Denmark,\\ Campusvej 55, DK-5230 Odense M, Denmark}

\author{Miguel Pato}
\affiliation{The Oskar Klein Centre for Cosmoparticle Physics, Department of Physics, Stockholm University, AlbaNova, SE-106 91 Stockholm, Sweden}

\date{\today}

\begin{abstract}
We introduce a new type of gamma-ray spectral feature, which we denominate gamma-ray triangle. This spectral feature arises in scenarios where dark matter self-annihilates via a chiral interaction into two Dirac fermions, which subsequently decay in flight into another fermion and a photon. The resulting photon spectrum resembles a sharp triangle and can be readily searched for in the gamma-ray sky. Using data from the Fermi-LAT and H.E.S.S.~instruments, we find no evidence for such spectral feature and therefore set strong upper bounds on the corresponding annihilation cross section. A concrete realization of a scenario yielding gamma-ray triangles consists of an asymmetric dark matter model where the dark matter particle carries lepton number. We show explicitly that this class of models can lead to intense gamma-ray spectral features, potentially at the reach of upcoming gamma-ray telescopes, opening a new window to explore asymmetric dark matter through indirect searches.
\end{abstract}

\maketitle

\section{Introduction}\label{sec:intro}
\par Cosmological and astrophysical observations, most recently enhanced by the input of the Planck satellite \cite{Ade:2013zuv}, have revealed that approximately 16\% of the matter content of the universe is in the form of baryons and 84\% in the form of a non-luminous component, dubbed dark matter. The same data also indicates that the primordial antibaryon abundance is negligible relative to the baryon abundance, whereas the dark antimatter abundance is currently unknown. While the particle physics properties of the baryonic content are by now very well understood, the nature of dark matter is largely unknown. A plausible assumption is that dark matter is composed of a new particle not contained in the Standard Model (SM). If this is the case, the new particle is likely to produce observable effects other than gravitational, an exciting possibility that has triggered an ambitious experimental program with three complementary strategies: direct detection, indirect detection and collider searches (for reviews, see Refs.~\cite{BertoneReview,BergstromReview,Feng:2010gw}).

\par The origin of the matter content of the universe, both baryonic and dark, remains as one of the most important open questions in cosmology. The known properties of the proton strongly suggest that the present population of baryonic matter is a result of an asymmetry in the number densities of baryons and antibaryons, dynamically generated after inflation (for a review, see Ref.~\cite{Dine:2003ax}). This mechanism, furthermore, can be implemented in simple particle physics models when the three Sakharov conditions \cite{Sakharov:1967dj} are simultaneously fulfilled (some renown realizations were proposed in Refs.~\cite{Rubakov:1996vz,Fukugita:1986hr,Affleck:1984fy}). On the other hand, the origin of the dark matter content is still widely debated. The most popular mechanism is freeze-out \cite{Zeldovich,Chiu:1966kg,Steigman:1979kw,Scherrer:1985zt}, which requires dark matter to be a weakly interacting massive particle (WIMP), namely a particle with mass in the GeV-TeV range and coupling to SM particles with a strength comparable to that of the electroweak force. The interactions which allow the dark matter freeze-out also lead to potentially observable signals in direct, indirect and collider experiments, thus providing avenues to test the WIMP hypothesis. In this framework, however, the similarity between the baryon and dark matter abundances turns out to be merely coincidental. 

\par An alternative dark matter production mechanism consists in the generation of an asymmetry in the number densities of dark matter particles and antiparticles at very early times, in complete analogy with the baryogenesis mechanism~\cite{Nussinov:1985xr}. In this class of models, commonly known as asymmetric dark matter models (for reviews, see Refs.~\cite{Zurek:2013wia,Petraki:2013wwa}), the dark matter particle transforms non-trivially under a conserved or approximately conserved global ``dark matter symmetry'' (analogous to the baryon symmetry) and there exists an interaction that permits the annihilation of dark matter particles and antiparticles. This scheme has the virtue that  the  baryon and dark matter densities have a related origin, hence their number densities today can be naturally of the same order, in agreement with observations. However, indirect signals from annihilation in asymmetric dark matter models are generically expected to be very suppressed: the dark matter particle-antiparticle annihilation occurs with very small rates due to the tiny relic density of dark matter antiparticles, and the particle-particle annihilation is forbidden by the dark matter number conservation.

\par In this paper, we identify a class of asymmetric dark matter models where annihilation signals can occur at a sizable rate. Under certain conditions, these signals arise in the form of sharp spectral features, thus opening the possibility of efficiently probing the scenario with indirect search experiments. Take the case of a dark matter particle charged under a conserved (or approximately conserved) global symmetry with the same charge as a given chiral fermion. Then, the self-annihilation of dark matter particles is allowed if the final state contains two chiral fermions (or two antifermions, depending on the dark matter charge). We consider a particular realization of this scenario where the dark matter number is identified with the lepton number and dark matter is stabilized by an additional symmetry. The simplest annihilation channel allowed by the lepton number conservation has a final state consisting of two neutrinos; limits on the annihilation cross section into monoenergetic neutrinos have been derived in \cite{Aartsen:2015xej,Adrian-Martinez:2015wey,Wendell:2014dka}.

\par We study instead final states which produce sharp spectral features in the gamma-ray energy spectrum. This can be realized with dark matter annihilations via a chiral interaction into two Dirac fermions, singlets under the SM gauge group and with the same lepton number as the dark matter particle, which then decay in flight into SM particles, such that in the whole annihilation process the total lepton number is preserved. In particular, the fermion singlet can decay into a photon and a neutrino (or antineutrino) with a sizable branching ratio. As we shall argue next, such dark matter cascade annihilation generates a characteristic gamma-ray spectrum resembling a triangle. This new signal adds to the list of sharp spectral features known to arise in particle physics scenarios (for a phenomenological analysis, see Ref.~\cite{Tang:2015meg}): gamma-ray lines \cite{Srednicki:1985sf,Rudaz:1986db,Bergstrom:1988fp}, internal electromagnetic bremsstrahlung \cite{Bergstrom:1989jr,Flores:1989ru,Bringmann:2007nk} and gamma-ray boxes \cite{Box2012}.

\section{Gamma-ray triangles}\label{sec:pheno}
\par For the sake of simplicity, we shall describe in this section the particular case of a dark matter particle $\chi$ self-annihilating into two intermediate fermionic states $\psi$, which then decay in flight into a standard neutrino and a photon: $\chi\chi \to \psi\psi \to 2\nu2\gamma$. Sec.~\ref{sec:model} is devoted to realizing this scenario with a simplified model of asymmetric dark matter. Clearly, the mass hierarchy must be such that $m_\chi \geq m_\psi \geq m_\nu\approx 0$. In the rest frame of the fermion $\psi$, the photon is monochromatic with energy $E_\gamma'=m_\psi/2$. Nevertheless, when boosted to the laboratory frame (where dark matter is essentially at rest), the photon energy depends on the emission angle and lies within the kinematic ends $E_\pm$: 
\begin{equation}\label{eq:ends}
E_\pm = \left(1\pm \sqrt{\dpc}\right)\, \frac{m_\chi}{2} \quad \mbox{with} \quad \delta_{ij}=1-\frac{m_i^2}{m_j^2} \, .
\end{equation}
For later convenience, we write down explicit expressions for the central energy $E_c\equiv(E_-+E_+)/2$ and the relative width $\Delta E/E_c\equiv (E_+-E_-)/E_c$:
\begin{equation}\label{eq:centralwidth}
\frac{E_c}{m_\chi} = \frac{1}{2} \quad ,\quad \frac{\Delta E}{E_c} = 2\sqrt{\dpc} \, .
\end{equation}

\par The above results follow directly from kinematic considerations and apply as well to the case of gamma-ray boxes. No information is given so-far regarding the shape of the spectrum between the edges $E_\pm$. This shape is defined by the angular distribution of the emitted photons that can be parametrized as 
\begin{equation}\label{eq:ftheta}
\frac{\D f}{\D \cos{\theta'}}=\frac{1}{2}\left( 1 + \alpha \cos{\theta'}\right) \, ,
\end{equation}
where $-1\leq \alpha \leq 1$ measures the spin polarization of the parent fermion $\psi$ and $\theta'$ is the angle between the photon momentum in the rest frame of $\psi$ and the momentum of  $\psi$ in the laboratory frame. This is a well-known result in the literature, e.g.~\cite{Raffelt:1996wa}. If $\psi$ is a Majorana fermion (or a scalar), there is no preferential emission direction, therefore $\alpha=0$ and we recover the case of gamma-ray boxes. If $\psi$ is instead a Dirac fermion, the spin of the particle defines a preferential direction: photons are emitted in the forward direction if $\alpha>0$ and in the backward direction if $\alpha<0$. The actual value of $\alpha$ is defined by the details of the particular model (for a worked-out example, see Sec.~\ref{sec:model}); in particular, if $\psi$ is emitted essentially at rest (i.e.~$\dpc\approx 0$) then $\alpha\approx 0$, while if $\psi$ is highly relativistic (i.e.~$\dpc\approx 1$) then $|\alpha|\approx1$. In the latter case, the photon is emitted essentially with a fixed polarization.
All in all,  it is necessary that the intermediate Dirac fermions  are  mostly produced in a state of fixed helicity $\pm1/2$. This cannot be realized if $\chi$ is a Majorana fermion or if $\chi$ and $\overline{\chi}$ are  thermal relics. On the other hand, in some classes of asymmetric dark matter scenarios, where either the particle or the antiparticle dark matter abundance is highly suppressed, the conditions to obtain this spectral feature are easily accommodated, as we shall see in a specific model realization.

Finally, convoluting the energy and angular emission spectrum of the photons and boosting to the laboratory frame we get the normalized photon spectrum
\begin{equation}\label{eq:dNdx}
\frac{\D N_\gamma}{\D x} = \frac{N_\gamma}{\dpc} \left( \sqrt{\dpc}-\alpha + 2\alpha x \right)
 \Theta\left(x-x_- \right)\Theta\left(x_+-x \right)
\end{equation}
with $x=E_\gamma/m_\chi$, $x_\pm=E_\pm/m_\chi$, $N_\gamma$ the number of photons emitted per annihilation (in our case $N_\gamma=2$) and $\Theta$ the Heaviside function. The formalism above holds for the decaying dark matter scenario $\chi \to \psi X \to \nu \gamma X$ with $N_\gamma=1$ and the replacement $m_\chi\mapsto m_\chi/2$ with the corresponding redefinition of $x$ and $\dpc$. In the case of a non-standard neutrino with a generic mass $m_\nu$ (which we will explore in a forthcoming publication \cite{NextPaper}), the same formalism can be applied with the replacement $m_\chi\mapsto \dnp m_\chi$ in Eqs.~\eqref{eq:ends} and \eqref{eq:centralwidth} and the corresponding redefinition of $x$ in Eq.~\eqref{eq:dNdx}, without redefining $\dpc$.

\begin{figure*}[htp]
\center
\includegraphics[width=.49\textwidth]{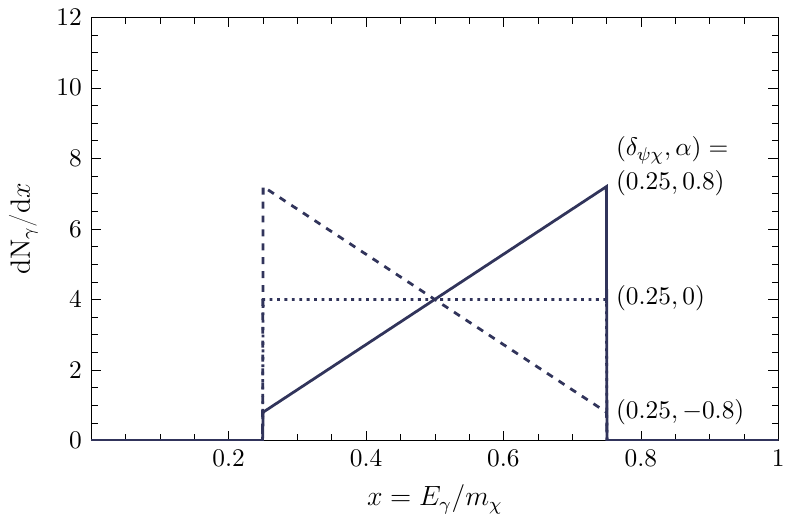}
\includegraphics[width=.49\textwidth]{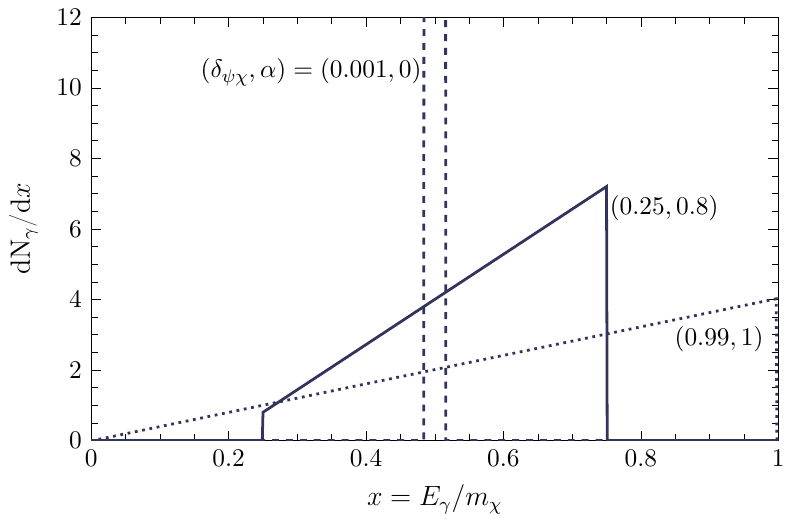}
\caption{\small The triangular photon spectrum of a cascade annihilation $\chi\chi \to \psi\psi \to 2\nu2\gamma$ with an intermediate fermionic state $\psi$. In the left panel, sample spectra are plotted for different values of $\alpha$, while fixing the mass splitting $\dpc= 0.25$. In the right panel, we show the effect of varying the mass splitting $\dpc$; for the extreme values of $\dpc$ we adopt the physical values of $\alpha$, i.e.~$\alpha=1$ for $\dpc= 0.99$ and $\alpha=0$ for $\dpc= 0.001$ (see text for a detailed discussion).
}\label{fig:spectra1}
\end{figure*}

\par As can be directly read from Eq.~\eqref{eq:dNdx}, the spectra have a constant linear slope $2N_\gamma\alpha/\dpc$ with sharp cutoffs at both kinematic ends. This is the triangular spectral feature we propose here in the context of dark matter searches and that we aim at constraining with current gamma-ray data. To be precise, the feature has a trapezoidal shape, but we classify it as triangular for simplicity of language. The typical shape of triangular spectra is illustrated in Fig.~\ref{fig:spectra1} for a baseline benchmark scenario with $\dpc=0.25$ (i.e.~$m_\psi/m_\chi= \sqrt{3}/2$) and $\alpha=0.8$, which is in line with the simplified model presented in Sec.~\ref{sec:model} (cf.~in particular Eq.~\eqref{eq:alpha}). The left and right panels show explicitly the effect of varying $\alpha$ and $\dpc$ alternately. Clearly, over a typical astrophysical background $\D \Phi_\text{bkg}/\D E_\gamma=A E_\gamma^{-p}$ with $2 \leq p \leq 3$, triangular spectra provide a sharp feature along the full kinematic range and in particular at the kinematic ends. Notice that both in the case of an up triangle ($\alpha>0$) and of a down triangle ($\alpha<0$) a spectral feature arises that can be cleanly looked for against a smooth power-law background. We gather from the right panel (or Eq.~\eqref{eq:centralwidth}) that, as the mass of the intermediate approaches the dark matter mass (i.e.~$\dpc \to 0$), the spectrum gets sharper and sharper until it effectively becomes a line. This behavior is very similar to the one observed for gamma-ray boxes in our previous works \cite{Box2012,Box2013}. The choice $\dpc=0.99$, instead of $\dpc=1$, is made to avoid a strictly massless intermediate fermionic state $\psi$ while having negligible effect on the photon spectrum itself. Following the discussion after Eq.~\eqref{eq:ftheta}, we adopt $\alpha=1$ ($\alpha=0$) for the benchmark with $\dpc=0.99$ ($\dpc=0.001$) to be consistent; notice however that the shape of these spectra would be very similar had we simply taken $\alpha=0.8$.

\par The spectral features introduced above produce a photon flux at Earth given by 
\begin{eqnarray}\label{eq:unconvolphiAnn}
\frac{\D \Phi_\text{ann}}{\D E_\gamma} &=& \frac{( \sigma v )_{0}^{2\gamma}}{8\pi m_\chi^2} \, \frac{\D N_{\gamma}}{\D E_{\gamma}} \, \bar{J}_{\text{ann}} \, ,\\
 \bar{J}_{\text{ann}} &\equiv&  \frac{J_\text{ann}}{\Delta\Omega} \equiv  \frac{1}{\Delta\Omega} \int_{\Delta \Omega}{\D\Omega \, \int_{\text{los}}{\D s \, \rho_{\text{dm}}^2 } }
\end{eqnarray}
for self-annihilating particles $\chi$ constituting most of the dark matter, where $( \sigma v )_0^{2\gamma}$ is the  annihilation cross section of the process $\chi\chi \to \psi\psi \to 2\nu2\gamma$, $\D N_{\gamma}/\D E_{\gamma}\equiv m_\chi^{-1} \D N_{\gamma}/\D x$ is the injection spectrum introduced in Eq.~\eqref{eq:dNdx}, $\bar{J}_{\text{ann}}$ ($J_\text{ann}$) is the averaged (actual) annihilation J-factor, $\Delta\Omega$ is the target field of view usually defined by a range of Galactic coordinates $(\ell,b)$, $s$ is the distance along the line of sight and $\rho_\text{dm}$ is the density of dark matter in the Galaxy. Notice that Eq.~\eqref{eq:unconvolphiAnn} is valid both for self-annihilating Majorana (symmetric) dark matter and for self-annihilating Dirac asymmetric dark matter as long as the self-annihilating Dirac fermion contributes the most of the dark matter budget (as opposed to the corresponding antifermion); the latter will be the case in the simplified model introduced in Sec.~\ref{sec:model}. For simplicity, and bearing in mind that the dark matter distribution in our Galaxy is a major uncertainty for indirect searches, we assume throughout an Einasto profile \cite{Navarro:2003ew,Merritt2006} with scale radius $r_s=20\,$kpc, slope parameter $\alpha_\text{Ein}=0.17$, local dark matter density $\rho_0\equiv\rho_\text{dm}(R_0)=0.4\,\textrm{GeV/cm}^{3}$ \cite{CatenaUllio2010,Weber:2009pt,Salucci:2010qr,2011MNRAS.414.2446M,Iocco2011,Nesti:2013uwa} and a distance of the Sun to the Galactic center $R_0=8.5\,$kpc \cite{Gillessen2009,Ando2011,Malkin2012,Reid2014}. The predicted flux in Eq.~\eqref{eq:unconvolphiAnn} can now be tested against gamma-ray data from Fermi-LAT and H.E.S.S..

\begin{figure}[htp!]
\center
\hspace{-0.15cm}\includegraphics[width=0.49\textwidth]{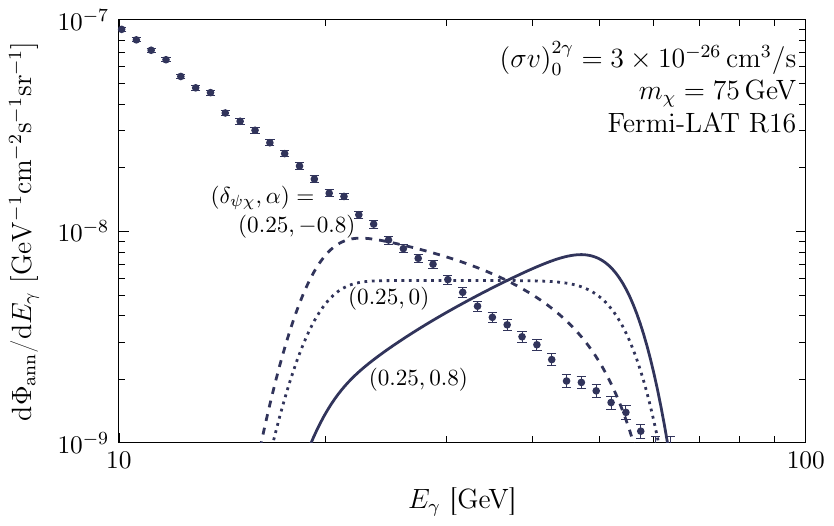}
\caption{\small The signal expected from triangular spectral features against current Fermi-LAT data. The signal plotted is produced by a dark matter model with $( \sigma v)_0^{2\gamma}=3\times10^{-26}\,\text{cm}^3/\text{s}$, $m_\chi=\unit[75]{GeV}$, $\dpc=0.25$ and $\alpha=0,\pm0.8$, taking into account the energy resolution of Fermi-LAT. Note that these features correspond to the spectra shown in the left panel of Fig.~\ref{fig:spectra1}. The Fermi-LAT data correspond to a circular region of radius $16^\circ$ around the Galactic centre (R16; cf.~Appendix \ref{app:data}). The cross section value $3\times10^{-26}\,\text{cm}^3/\text{s}$ used here is solely for illustrative purposes and has no physical relevance, since we are not considering thermal relics but asymmetric dark matter models.
}
\label{fig:specdata}
\end{figure}

\begin{figure*}[htp]
\center
\includegraphics[width=.49\textwidth]{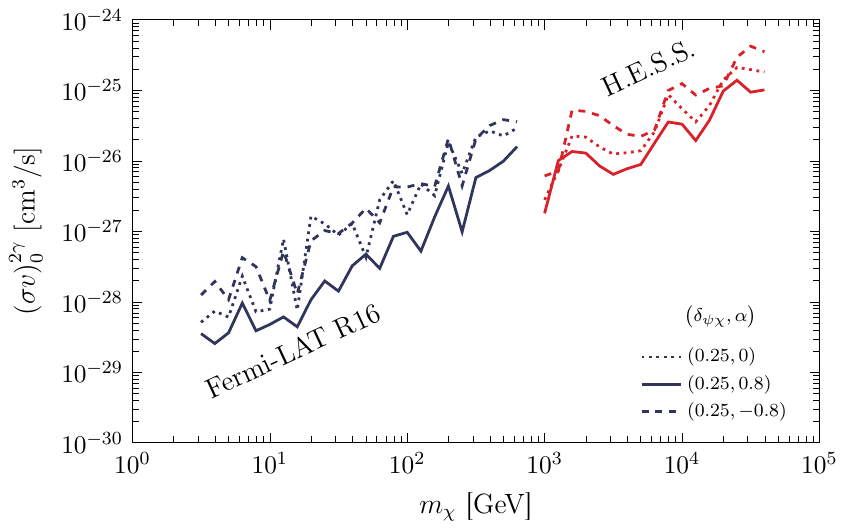}
\includegraphics[width=.49\textwidth]{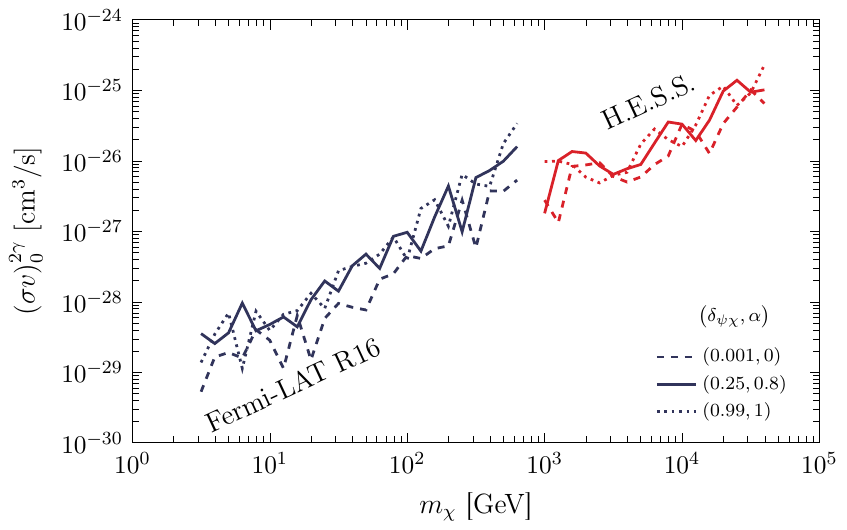}
\caption{\small The one-sided 95\% confidence level upper limits on the annihilation cross section $(\sigma v )_{0}^{2\gamma}$ for triangular features of fixed width and different slopes (left) and different widths (right). The benchmark configurations shown here mimic the ones in the left and right panels of Fig.~\ref{fig:spectra1}. The black and red lines correspond to the bounds obtained using Fermi-LAT R16 and H.E.S.S.~data sets, respectively (cf.~Appendix \ref{app:data} for further details on data treatment).
}\label{fig:upperlimits}
\end{figure*}

\par In order to illustrate the derivation of our bounds on triangular features, we plot in Fig.~\ref{fig:specdata} the signal expected for three dark matter models with a nominal cross section against a particular Fermi-LAT data set. The signal is appropriately convoluted with the energy resolution of the Fermi-LAT instrument. The triangular spectral features differ starkly from a power-law gamma-ray flux in the cases where $\alpha\geq0$ and can thus be strongly constrained with present data already. This is less evident when $\alpha<0$; however, let us notice that in this case the signal is still harder than a soft power law as seen in Fig.~\ref{fig:specdata} (cf.~also Eq.~\eqref{eq:dNdx}). Clearly, the three dark matter benchmarks with the particular choice of cross section are strongly excluded by Fermi-LAT data.

\par We now set out to derive precise upper limits on the annihilation cross section using a profile likelihood analysis for different data sets and different parameter configurations. For a detailed account of our data treatment and derivation of limits, please refer to Appendix \ref{app:data}. Fig.~\ref{fig:upperlimits} shows the one-sided 95\% confidence level (CL) upper limits on the annihilation cross section $(\sigma v )_{0}^{2\gamma}$ for the triangular spectra presented in Fig.~\ref{fig:spectra1}. The complementarity between Fermi-LAT and H.E.S.S.~is immediately apparent and provides effective bounds on triangular spectral features over almost five orders of magnitude in energy (or, equivalently, in dark matter mass) without gaps. This is a remarkable achievement for gamma-ray telescopes and it carries crucial importance for indirect dark matter searches. The jagged aspect of the limits is due to the expected statistical fluctuations in the gamma-ray data. The left panel of Fig.~\ref{fig:upperlimits} shows the cases of triangles of different slopes but constant width (cf.~left panel of Fig.~\ref{fig:spectra1}), while the right panel focuses on up triangles of different widths (cf.~right panel of Fig.~\ref{fig:spectra1}). Intuitively, one would expect that the harder the signal, the stronger the limits when searching against a power-law background. In fact, at face value, the strongest limit is obtained for the narrow, line-like triangle corresponding to the configuration $(\dpc,\alpha)=(0.001,0)$. Notice however that signals less sharp but wider than a line extend to higher energies where the background is smaller, therefore resulting in rather strong limits, as also observed for box spectra \cite{Box2012,2015JCAP...09..048I}. A similar situation holds for down triangles ($\alpha<0$), for which the limits in the left panel of Fig.~\ref{fig:upperlimits} can be improved by sliding down the energy window, cf.~Appendix \ref{app:data}. Overall, the upper limits range from $(\sigma v )_{0}^{2\gamma} \approx 10^{-29}\,\text{cm}^3/\text{s}$ at dark matter masses of a few GeV up to $( \sigma v )_{0}^{2\gamma} \approx 10^{-25}\,\text{cm}^3/\text{s}$ at several tens of TeV masses. 

\par One should note that the intensity of the sharp spectral feature produced in the cascade annihilation $\chi\chi\rightarrow \psi\psi$ followed by $\psi\rightarrow \nu\gamma$ is neither suppressed by a factor $\alpha_{\rm em}^2/(4\pi)^2$, as is the case of the gamma-ray line, nor by $\alpha_{\rm em}/\pi$, as is the case of the internal electromagnetic bremsstrahlung. Therefore, the non-observation of sharp features in the gamma-ray energy spectrum provides strong limits on this class of asymmetric dark matter models, as we will show in the next section with a concrete example.

\section{A simplified model of asymmetric dark matter}\label{sec:model}
\par We introduce a simplified  model which realizes the gamma-ray spectral features discussed in the previous section. The model is characterized by two Dirac spinors, $\chi$ and $N_D$, and one real scalar field $S$. The new fields are singlets of the SM gauge group and carry a global $U(1)_{\rm X}$ charge. In addition, we impose a $Z_2$ discrete symmetry to the overall Lagrangian, under which only $\chi$ and $S$ transform non-trivially. The particle content of the theory and the charge assignments of the fields are summarized in Tab.~\ref{tab:FieldAssign}. We define a non-trivial transformation rule of the SM left-handed (right-handed) leptons, $L_\alpha$ ($e_{R\alpha}$) for $\alpha=e,\mu,\tau$, under the global symmetry, therefore $U(1)_{\rm X}$ can be identified with the total lepton number.

\begin{table}
\centering
\begin{tabular}{|c||c|c|c|c||c|}
\hline 
\rule[0.15in]{0cm}{0cm}{\tt Field} & $L_{\alpha}$ & $e_{R\alpha}$ & $N_D$ & $\chi$ & $S$  \\
\hline
$U(1)_{\rm X}$ &$1$ & $1$ & $1$ & $1$ & $0$ 
\\\hline 
$Z_2$ & $1$ & $1$ & $1$  & $-1 $& $-1$ 
\\\hline
\end{tabular}
\caption{\small Particle content and charge assignments of our asymmetric dark matter simplified model.}\label{tab:FieldAssign}
\end{table}

\par The Lagrangian of the model is the following:
\begin{eqnarray}
\mathcal{L}  & = & \mathcal{L}_{\rm SM}  +  \frac 1 2 \partial_\mu S  \partial^\mu S +  i \overline{\chi}  \slashed{\partial}  \chi +  i  \overline{N}_D  \slashed{\partial}  N_D  \nonumber\\
&& -\frac 1 2 \mu_S^2  S^2  -  m_\chi \overline{\chi} \chi -  m_N \overline{N}_D N_D   \nonumber\\
&& - \left( \lambda_\alpha  \overline{L}_\alpha  N_D  H^c + f  \overline{N}_D P_{L,R} \chi  S   +  {\rm h.c.} \right) \nonumber\\
&& - V(H, S) \, ,\label{Lagr1}
\end{eqnarray}
where $\mathcal{L}_{\rm SM}$ denotes the SM Lagrangian, $H$ is the SM Higgs doublet,  $H^c\equiv  \epsilon H^*$ is the charge-conjugated field, and $L_\alpha\equiv (\nu_{\alpha L}, \ell_{\alpha})$. Notice that in the third line of Eq.~\eqref{Lagr1} we consider a chiral Yukawa interaction term between $N_D$, $\chi$ and $S$, without specifying the chiral projector for the moment. As we shall see below, this choice determines if the photon spectrum resulting from the  decay of the intermediate fermion is peaked at high or low energies. In what follows we shall work in the basis where the coupling constant $f$ is real and positive. 

\par The scalar quartic potential $V(H, S)$ is given by
\begin{equation}
	V(H, S) = \lambda_{HS}\left(H^\dagger H \right) S^2 + \lambda_S  S^4\,.
\end{equation}
 We will assume parameters of the Lagrangian such that $S$ does not acquire a vacuum expectation value (vev) and, therefore, the $Z_2$ symmetry is preserved. After electroweak symmetry breaking, the mass of the scalar field $S$ reads
\begin{equation}
m_S =  \sqrt{\mu_S^2 +  \lambda_{HS}   v_H^2 } \, ,
\end{equation}
$v_H\simeq 246$ GeV being the SM Higgs vev.

\par We assume that the Dirac fermion $\chi$ is the lightest particle in the $Z_2$-odd sector, therefore it is absolutely stable and represents a dark matter candidate. In this scenario, dark matter annihilates into a pair of Dirac fermions $N_D$, which subsequently decay in flight into SM particles, being the lepton charge preserved throughout the whole process. In particular, as shown below, the radiative decay of $N_D$, due to the mixing with the light active neutrinos, produces a photon spectrum with the triangular shape described in Sec.~\ref{sec:pheno}.

\par We assume here that the dark matter abundance observed today has an origin analogous to the baryonic (visible) matter, that is, an asymmetry between $\chi$ and $\overline{\chi}$ was produced at a certain time in the early universe. We do not specify the mechanism at the origin of such asymmetry, but impose that almost all the dark matter density today is made of the particle $\chi$, while the corresponding antiparticle abundance is negligible (concrete frameworks of dark matter production were discussed in \cite{Iminniyaz:2011yp,Falkowski:2011xh}). The analysis is equivalent if one assumes instead that $\overline{\chi}$ is the dominant dark matter component.
 
\par The present-day annihilation of a pair of $\chi$ into two (on-shell) intermediate fermions $N_D$ proceeds via $s$-wave and the corresponding cross section reads
\begin{equation}
	(\sigma v)_0 = \frac{f^4}{64 \pi m_\chi^2} \frac{\left(1+ \dNc \right)\sqrt{\dNc}}{\left(1 - \dSc + \dNc \right)^2} \, ,\label{csunpol}
\end{equation}
where the mass splittings $\dNc$ and $\dSc$ follow the definition in Eq.~\eqref{eq:ends}. Taking as benchmark values $m_\chi = 50\,\text{GeV}$ and  $m_S = 100\,\text{GeV}$, we have
\begin{equation}
	(\sigma v)_0 \approx 1.86 \times 10^{-24} f^4 \,\text{cm}^3/\text{s}
\end{equation}
for $m_N \ll m_\chi$ and
\begin{equation}
        (\sigma v)_0 \approx 1.67 \times 10^{-24} f^4 \,\text{cm}^3/\text{s}
\end{equation}
for $m_N = 20\,$GeV. Hence, for a coupling $f\approx\mathcal{O}(1)$, we obtain annihilation cross sections much larger than the typical present-day value for $s$-wave self-annihilating thermal relics, $(\sigma v )_{\rm th,0} \approx 3\times 10^{-26}\,\text{cm}^3/\text{s}$.

\par The Dirac fermions in the final state  decay into SM particles through Higgs-mediated, charged-current (CC) and neutral-current (NC) interactions. The latter are generated after electroweak symmetry breaking  due to the mixing between active neutrinos and $N_D$. The interaction terms are the following:
\begin{eqnarray}
 \mathcal{L}_{CC}^N &=& -\frac{g}{2 \sqrt{2}}
\overline{\ell}_{\alpha}\gamma_{\mu}\Theta_{\alpha}(1 - \gamma_5)N_{D}W^{\mu}+ {\rm h.c.} \, ,
\label{NCC}\\
 \mathcal{L}_{NC}^N &=& -\frac{g}{4 c_W}\overline{\nu}_{\alpha L}\gamma_{\mu}\Theta_{\alpha}(1 - \gamma_5)N_{D}Z^{\mu}+ {\rm h.c.} \, ,
\label{NNC}\\
\mathcal{L}_{H}^N &=& -\frac{g m_{N}}{4 m_{W}}\overline{\nu}_{\alpha L}\Theta_{\alpha}(1 + \gamma_5)N_{D}h + {\rm h.c.} \, ,
\label{NH}
\end{eqnarray}
where $W$ and $Z$ denote the electroweak gauge bosons, $h$ is the Higgs boson, $g$ is the weak coupling, $c_W$ is the cosine of the weak mixing angle $\theta_W$ and $\Theta_\alpha\approx \lambda_\alpha v_H/m_N \ll 1$ is the mixing between SM neutrinos and $N_D$. The  mixing $\Theta_\alpha$ is strongly constrained by both direct and indirect searches of sterile neutrinos (see e.g.~\cite{Kusenko:2009up, Atre:2009rg, Deppisch:2015qwa}). Notice that the flavor structure of the coupling is fully determined by neutrino oscillation data if we consider a generalization of our model where we introduce a low-scale type I seesaw scenario with two Majorana neutrinos that form a pseudo-Dirac pair, which can be identified with the field $N_D$ (see e.g.~\cite{JosseMichaux:2011ba, JosseMichaux:2012wj}). In this case, the total lepton charge symmetry is softly broken and all lepton number violating processes are effectively suppressed by the light neutrino mass scale \cite{delAguila:2008hw,Gavela:2009cd,Ibarra:2010xw,Ibarra:2011xn,Dinh:2012bp,Lopez-Pavon:2015cga}.

\begin{figure}[t!]
\center
\hspace{-0.55cm}\includegraphics[width=.48\textwidth]{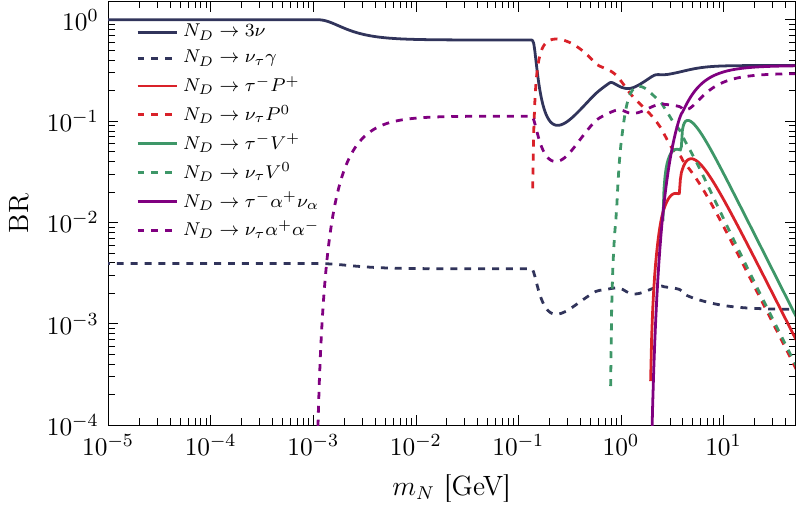}\\
\hspace{-0.55cm}\includegraphics[width=.48\textwidth]{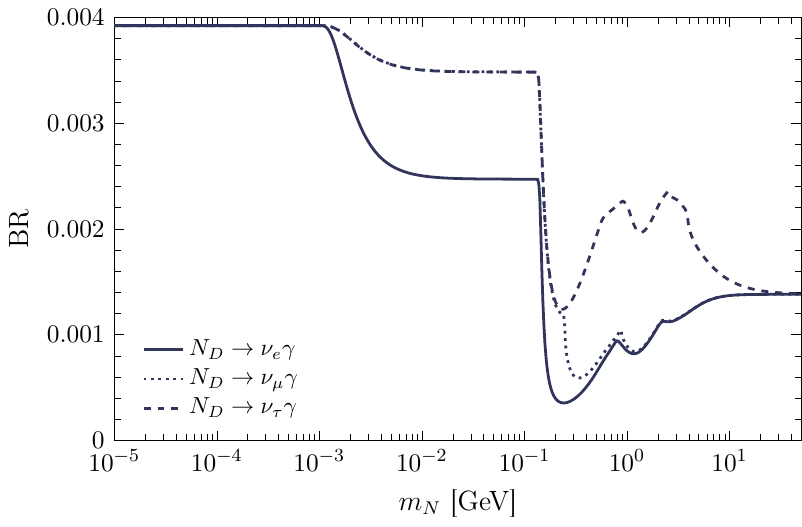}
\caption{\small The branching ratios of $N_D$ (semi-)leptonic decays into SM particles. In the upper panel, $P^{0,+}$ ($V^{0,+}$) denote the pseudo-scalar (vector) mesons which are kinematically accessible (see e.g.~\cite{Atre:2009rg}) and it is assumed that $N_D$ couples only to the third lepton family. In the lower panel, the branching ratio of $N_D$ radiative decay is shown for a coupling to the first, second and third lepton families.
} \label{fig:BR}
\end{figure}

\par The upper panel of Fig.~\ref{fig:BR} shows the branching ratio of $N_D$ decays into SM particles, that is, into pseudo-scalar/vector mesons and leptons. We assume, for the sake of discussion, that $N_D$ couples only to one lepton flavor, in this case to the third family. Clearly, for $m_N\lesssim 1\,\text{MeV}$ the dominant decay channel is the one with three neutrinos in the final state. The decay rate of $N_D$ is in this case~\cite{Atre:2009rg}
\begin{equation}
	\Gamma(N_D \to \nu_\alpha \nu_\beta  \overline{\nu}_\beta) = \frac{G_F^2}{96 \pi^3} |\Theta_\alpha|^2 m_N^5 \, ,
\end{equation}
where $G_F\equiv\sqrt{2}g^2/(8m_W^2)$ is the Fermi constant. The second open decay channel in the low-mass regime is via the emission of a photon and a (left-handed) SM neutrino\footnote{In the case of a dark matter abundance dominated by the $\overline{\chi}$ component, the photon spectrum arises from the cascade process $\overline{\chi}\,\overline{\chi} \to  \overline{N}_D \overline{N}_D$,  $\overline{N}_D\to \overline{\nu}_\alpha \gamma$, where $\overline{\nu}_\alpha$ is a positive-helicity state.}, which arises at one-loop order. The corresponding decay rate is \cite{Pal:1981rm}
\begin{equation}
	\Gamma(N_D \to \nu_\alpha \gamma) =\frac{9 \alpha_{\rm em} G_F^2}{512 \pi^4} |\Theta_\alpha|^2 m_N^5 \, ,
\end{equation}
where $\alpha_{\rm em}$ is the fine-structure constant. We report in the lower panel of Fig.~\ref{fig:BR} the branching ratio of $N_D$ radiative decay for $m_N\leq 50\,\text{GeV}$ in the case $N_D$ couples exclusively to the electron, muon and tau lepton flavors. The asymptotic values are
\begin{equation}
	{\rm BR}(N_D \to \nu_\alpha \gamma) \approx 0.0039
\end{equation}
for $m_N\lesssim 1$\,MeV and
\begin{equation}
	{\rm BR}(N_D \to \nu_\alpha \gamma) \approx 0.0014
\end{equation}
for $m_\tau \ll m_N\ll 100\,\text{GeV}$. For values of $m_N\gg 50$ GeV, the total width of $N_D$ is dominated by the decay into $W$, $Z$ and the Higgs boson. Accordingly, the branching ratio $N_D\to \nu_\alpha \gamma$ is highly suppressed. In this mass regime, nonetheless, the decay of $N_D$ generates a flux of continuum gamma-rays, antimatter particles and neutrinos which might also be at the reach of experiments. 
We have checked this statement explicitly for the three benchmarks used in the right panel of Fig.~\ref{fig:spectra1}  by computing numerically the photon spectrum arising from the cascade decays $N_D\to \tau^-\pi^+$ and $N_D\to\nu_\tau\pi^0$ using \textsc{Pythia} 6.4 \cite{Sjostrand:2006za}. We find that for $(\delta_{N\chi},\alpha)\simeq(0,0),\,(0.25,0.8),\,(0.99,1)$ the spectral feature dominates over the secondary production of photons in the mass range $m_\chi\gtrsim 5,\,15,\,100\,$GeV, respectively. The case of $\alpha<0$ is more easily dominated by the secondary production.

\par The photon spectrum of the full process $\chi\chi \to N_D N_D \to 2\nu_\alpha 2\gamma$ features a central energy $E_c/m_\chi = 1/2$ and a relative width $\Delta E/E_c = 2\sqrt{\dNc}$, cf.~Eq.~\eqref{eq:centralwidth}. The spin polarization $\alpha$, which characterizes the photon distribution in Eq.~\eqref{eq:ftheta} and the slope of the photon spectrum in Eq.~\eqref{eq:dNdx}, only depends on the masses of the dark matter particle and the intermediate fermion (cf.~Appendix \ref{app:spin}). The result  in this simplified model with a scalar mediator is
\begin{equation}\label{eq:alpha}
	\alpha=\pm  \frac{2 \sqrt{\dNc}}{1 + \dNc} \, ,
\end{equation}
where the plus (minus) sign is obtained from the right-handed (left-handed) chiral projector $P_R$ ($P_L$) in the interaction Lagrangian of Eq.~\eqref{Lagr1}. For a non-relativistic $N_D$ (i.e.~$\dNc\approx 0$), $\alpha\approx 0$ and the resulting photon spectrum is a box, as for the radiative decay of a Majorana fermion. This is expected because in this limit  there is an equal probability of emitting a photon in the forward and backward directions, cf.~Eq.~\eqref{prpol}. Notice, however, that for highly degenerate $\chi$ and $N_D$ the photon spectrum is effectively a line (at $m_\chi/2$) and is hardly sensitive to the actual value of $\alpha$. In contrast, for a fully relativistic intermediate fermion (i.e.~$\dNc\approx 1$), the spin polarization is almost maximal, $|\alpha|\approx 1$, and the photon spectrum is extended.

\par
\begin{figure}[tp]
\center
\hspace{-0.15cm}\includegraphics[width=.49\textwidth]{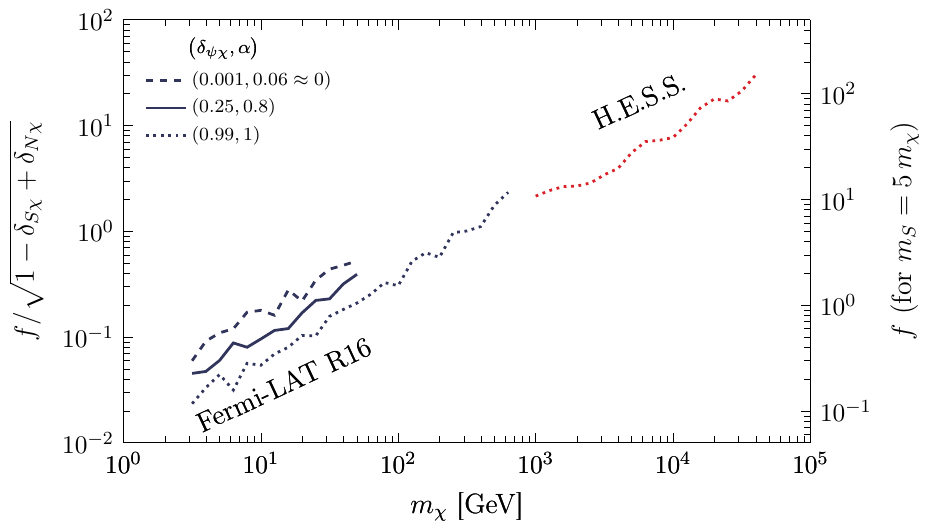}
\caption{\small The one-sided 95\% confidence level upper limits on the coupling $f$ for the three benchmarks of Fig.~\ref{fig:upperlimits} (right). The limits are shown for a generic mass splitting $\delta_{S \chi}$ (left axis) and for $m_S=5~m_\chi$  (right axis), where in the latter case $f$ becomes non-perturbative for $m_\chi\gtrsim 1.3$ TeV.
}\label{fig:flimits}
\end{figure}

 We show in Fig.~\ref{fig:flimits} the upper bound on the coupling $f$ imposed by our analysis of Fermi-LAT and H.E.S.S.~data. We assume $N_D$ equally coupled to each neutrino flavor. The quantity reported on the left vertical axis corresponds to the combination of $f$, $\delta_{N\chi}$  and $\delta_{S\chi}$ that enters in the expression of the annihilation cross section, Eq.~\eqref{csunpol}, whereas on the right axis the range of $f$ is displayed for $m_S= 5~m_\chi$ ($\delta_{S\chi}=-24$). We use the constraints corresponding to  the three benchmark points ($\delta_{N\chi}, \alpha$) in the right panel of Fig.~\ref{fig:upperlimits}, which fulfill Eq.~\eqref{eq:alpha}. Note that the cross section $(\sigma v)_0^{2\gamma}$ in Sec.~\ref{sec:pheno} and Fig.~\ref{fig:upperlimits} corresponds to $(\sigma v)_0^{2\gamma}\equiv (\sigma v)_0 \sum_\alpha{ \text{BR}(N_D\to\nu_\alpha \gamma) }$ in the notation of the present section. The range of $m_\chi$ to which the constraints apply strictly depends on the value of $\delta_{N\chi}$. In fact, for each curve reported in Fig.~\ref{fig:flimits}, that is, for each choice of $\delta_{N\chi}$ (and $\alpha$) we impose $m_N\leq 50\,$GeV. As discussed above, for larger masses, the $N_D$ decay channels into electroweak gauge bosons become kinematically allowed, thus strongly suppressing $\text{BR}(N_D\to\nu_\alpha \gamma)$. This in turn translates into an upper limit of the dark matter mass which leads to triangular gamma-ray spectral features in our model, namely $m_\chi\lesssim 50\,\text{GeV}/\sqrt{1-\delta_{N\chi}}$. Therefore, we have $m_\chi \lesssim 50\,(58)\,$GeV for $\delta_{N\chi} \approx 0\,(0.25)$, which is within the sensitivity of Fermi-LAT, cf.~dashed (solid) line in Fig.~\ref{fig:flimits}. Conversely, for $\delta_{N\chi}\approx 1$, any value of $m_\chi$ is viable in the model and both Fermi-LAT and  H.E.S.S.~constraints apply, cf.~black and red dotted lines. Notice that in the case $m_S =5~m_\chi$ shown in the plot, the coupling $f$ (to be read on right vertical axis) becomes non-perturbative, $f\gtrsim 4\pi$, for $m_\chi\gtrsim 1.3 $ TeV. The relative mass splitting $\delta_{N\chi}$ not only determines the gamma-ray spectral features in our scenario, but also affects the magnitude of the annihilation cross section, that is suppressed as $( \sigma v )_0^{2\gamma} \propto ( \sigma v )_0 \propto \sqrt{\delta_{N\chi}} \propto |\alpha|$ for $\sqrt{\delta_{N\chi}}\approx 0$ (cf.~Eqs.~\eqref{csunpol} and \eqref{eq:alpha}). For this reason, the strongest constraints in the low mass regime are set by Fermi-LAT data on wide triangles, i.e.~$\delta_{N\chi}\approx 1$. Conversely, the smaller $\delta_{N\chi}$, the weaker the bound on $f$.

\section{Conclusions}\label{sec:conc}
\par We introduced here a novel class of spectral features, which we denominate gamma-ray triangles, with important implications in the search for indirect signatures of dark matter. Gamma-ray triangles arise naturally in models where dark matter self-annihilates via a chiral interaction into two Dirac fermions, which subsequently decay in flight into another fermion and a photon. This scheme, while not applicable to standard thermal relics, can be easily embedded in some classes of asymmetric dark matter models, thus providing a possible hint of this class of candidates in indirect searches. The latest gamma-ray observations from Fermi-LAT and H.E.S.S.~strongly constrain gamma-ray triangles down to annihilation cross sections as low as $10^{-29}\,\text{cm}^3/\text{s}$. We illustrated the power of such bounds by constructing an explicit asymmetric dark matter setup and by singling out the large regions of the parameter space already ruled out. This points towards a promising avenue to look for a specific class of asymmetric dark matter models in a very efficient manner with gamma-ray observations, supplying complementary information to existing strategies in cosmology, direct searches and colliders. While the future observation of a triangular spectral feature would provide a remarkable hint of 
asymmetric dark matter and motivate a shift away from the WIMP paradigm, its non-detection would be instrumental in ruling out numerous asymmetric dark matter candidates. Either way, upcoming gamma-ray instruments, namely the Cherenkov Telescope Array, will have a decisive role in shaping our understanding of the nature of dark matter.
\\

\section*{Acknowledgements}
We would like to thank Tomer Volansky for discussions and especially Xiaoyuan Huang for his kind help in retrieving Fermi-LAT data and for useful input. This research was supported by the DFG cluster of excellence ``Origin and Structure of the Universe''. The CP$^3$-Origins center is partially funded by the Danish National Research Foundation, grant number DNRF90. M.~P.~acknowledges the support from Wenner-Gren Stiftelserna in Stockholm.

\appendix

\section{Data treatment}\label{app:data}
\par We use five sets of recent gamma-ray observations:
\begin{itemize}

\item {\bf Fermi-LAT 2011} \cite{fermilatsite,VertongenWeniger}. The Galactic centre region as defined in Ref.~\cite{VertongenWeniger} is used corresponding to $\bar{J}_{\text{ann}}=1.11 \times 10^{23}\,\text{GeV}^2/\text{cm}^5$. The data extends along the energy range $E_\gamma=1-300\,\text{GeV}$ with a mean exposure of $7.9\times 10^{10}\,\text{cm}^2\,\text{s}\,\text{sr}$ corresponding to $2.3\,\text{yr}$ of data taking between 2008 and 2011. For this data set, the Fermi-LAT energy resolution is modeled according to the Pass7\_V15 performance in Ref.~\cite{fermilatsite2}, featuring typical values $6.8-11.5\%$.

\item {\bf Fermi-LAT 2015} \cite{fermilatdata1,fermilatdata2}. We make use of the public Fermi-LAT gamma-ray data \cite{fermilatdata1} across energies $E_\gamma=1-500\,\text{GeV}$ between August 4, 2008 (MET 239557417) and Aug 5, 2014 (MET 428889603)\footnote{We would like to thank Xiaoyuan Huang for kindly providing access to this data set through the Fermi Science Tools.}. Selecting events from the PASS 8 SOURCE event class, standard diffuse analysis cuts are applied, including zenith angle $<90^\circ$ and the quality cut filter ``DATA-QUAL = 1, LAT-CONFIG = 1". The Fermi Science Tools (v10r0p5) \cite{fermilatdata2} are then used to calculate the exposure map. The data sets used in our work correspond to three distinct fields of view: a squared $2^\circ \times 2^\circ$ region around the Galactic centre (2x2), a circular $3^\circ$ region around the Galactic centre (R3) and a circular $16^\circ$ region around the Galactic centre with $|b|\leq5^\circ$ and $|l|\geq6^\circ$ masked out (R16). The first region is the one used in Ref.~\cite{2015JCAP...09..048I}, while the two other regions are inspired by the Fermi-LAT analysis \cite{Ackermann:2013uma}. The annihilation J-factors for the regions 2x2, R3 and R16 read $\bar{J}_{\text{ann}}=8.36\times 10^{24}\,\text{GeV}^2/\text{cm}^5$, $\bar{J}_{\text{ann}}=3.51\times 10^{24}\,\text{GeV}^2/\text{cm}^5$ and $\bar{J}_{\text{ann}}=5.12\times 10^{23}\,\text{GeV}^2/\text{cm}^5$. The mean exposures of the data are $2.94\times 10^{8}\,\text{cm}^2\,\text{s}\,\text{sr}$, 2.18$\times 10^{9}\,\text{cm}^2\,\text{s}\,\text{sr}$ and $4.69\times 10^{10}\,\text{cm}^2\,\text{s}\,\text{sr}$ for 2x2, R3 and R16, respectively. The Fermi-LAT energy resolution is modeled according to the Pass8\_R2\_V6 performance in Ref.~\cite{fermilatsite2}, featuring typical values $6.2-27.7\%$.

\item {\bf H.E.S.S.~2013} \cite{2013PhRvL.110d1301A}. Here we take the central Galactic halo defined by a circular $1^\circ$ around the Galactic centre with $|b|>0.3^\circ$, which corresponds to $\bar{J}_{\text{ann}}=7.78\times 10^{24}\,\text{GeV}^2/\text{cm}^5$. The data extends along the energy range $E_\gamma=500\,\text{GeV}-25\,\text{TeV}$ with a mean exposure of $2.20\times 10^{11}\,\text{cm}^2\,\text{s}\,\text{sr}$ corresponding to $112\,\text{h}$ of live time between 2004 and 2008. We assume an H.E.S.S.~energy resolution varying log-linearly with energy from $17\%$ at $500\,$GeV down to 11\% at $10\,$TeV, in line with the figures quoted in Ref.~\cite{2013PhRvL.110d1301A}.

\end{itemize}

\par For a given model configuration $(m_\chi,\dpc,\alpha)$ and each data set above, we perform a profile likelihood analysis \cite{Rolke:2004mj,VertongenWeniger,Bringmann:2011ye,Bringmannetal,Weniger:2012tx} with a model consisting of a background $\D \Phi_\text{bkg}/\D E_\gamma$ parametrized by a generic parameter vector $\bm p$ (two-parameter power law for Fermi-LAT or seven-parameter modulated power law for H.E.S.S.~\cite{2013PhRvL.110d1301A}) and a dark matter signal $\D \Phi_\text{ann}/\D E_\gamma$ (with strength given by $\mathcal{S}\equiv ( \sigma v )_0^{2\gamma}$, cf.~Eq.~\eqref{eq:unconvolphiAnn}). In the case of Fermi-LAT data we apply sliding energy windows $[\bar{E}/\sqrt{\epsilon},\bar{E}\sqrt{\epsilon}]$ with $\bar{E}=E_+$ and $\epsilon=1.5, 2.0, 2.3$, which correspond to 2$\sigma$ energy intervals for instruments with energy resolution ranging from 10 to 20\%, while for H.E.S.S.~we use the full energy range following Ref.~\cite{2013PhRvL.110d1301A}. For each energy bin $i$ with observed counts $n_{\text{obs}}^i$, the expected number of counts $n_{\text{exp}}^i({\bm p},\mathcal{S})$ is obtained by convoluting the expected flux $\D \Phi_\text{tot}/\D E_\gamma= \D \Phi_\text{bkg}/\D E_\gamma + \D \Phi_\text{ann}/\D E_\gamma$ with the energy resolution $\sigma_E$ and exposure $\mathcal{E}$ (in $\text{cm}^2\,\text{s}\,\text{sr}$) over the bin size. The likelihood then follows as a product of Poissonian probabilities over the bins inside the sliding energy window (for Fermi-LAT) or over the entire energy range (for H.E.S.S.),
\begin{equation}\label{eq:Lpoisson}
\mathcal{L}=\prod_i{P\left(n_{\text{obs}}^i | n_{\text{exp}}^i\right)} = \prod_i{  \frac{ (n_{\text{exp}}^i)^{n_{\text{obs}}^i}  \exp(-n_{\text{exp}}^i) } {n_{\text{obs}}^i!} }
\end{equation}
or 
\begin{equation}\label{eq:lnLpoisson}
\ln\mathcal{L}=\sum_i{  n_{\text{obs}}^i \ln n_{\text{exp}}^i - n_{\text{exp}}^i } \, ,
\end{equation}
where in the last expression we have dropped the term $-\ln (n_{\text{obs}}^i!)$, which is independent of ${\bm p}$ and $\mathcal{S}$ and therefore irrelevant for maximizing the likelihood. The quantity $-2\ln\mathcal{L}$ is then minimized over $\bm p$ for each value of $\mathcal{S}$, yielding the profile likelihood $-2\ln\mathcal{L}_{\text{prof}}(\mathcal{S})$. The overall minimum (best fit) is denoted $-2\ln\mathcal{L}_{\text{bf}}$ corresponding to the parameter set $({\bm p}_\text{bf},\mathcal{S}_\text{bf})$. The one-sided 95\% CL upper limit on signal strength is the value $\mathcal{S}_\text{ul}>\mathcal{S}_\text{bf}$ such that $-2\ln\mathcal{L}_{\text{prof}}(\mathcal{S}_\text{ul})=-2\ln\mathcal{L}_\text{bf}+2.71$ (see e.g.~\cite{1997sda..book.....C}). Finally, we assess the significance of a potential signal with the help of the usual test statistics
\begin{equation}\label{eq:TS}
TS=-2\left( \ln\mathcal{L}_\text{bf}^0 - \ln\mathcal{L}_\text{bf} \right) \, ,
\end{equation}
where $\mathcal{L}_\text{bf}^0=\mathcal{L}_{\text{prof}}(\mathcal{S}=0)$ is the maximum likelihood with no signal.

\par To avoid cluttering, the main results of our analysis, shown in Fig.~\ref{fig:upperlimits}, are conveyed in terms of upper limits on $(\sigma v )_0^{2\gamma}$ solely for the Fermi-LAT 2015 R16 region using a sliding window centered at $\bar{E}=E_+$ with width $\epsilon=2.0$ and for H.E.S.S.~2013 using the same exact procedure as in Ref.~\cite{2013PhRvL.110d1301A}. We have nevertheless tested and confirmed the robustness of our results. In particular, the use of the different Fermi-LAT data sets described above leads to upper limits equivalent to the ones of Fermi-LAT 2015 R16 region but weaker up to a factor of $\sim50$. Overall, the effect of narrowing the window width to $\epsilon=1.5$ or enlarging it to $\epsilon=2.3$ amounts to a mean factor of $\sim4$ in the upper limits. For the case of the down triangle ($\alpha<0$), an improvement of up to a factor of $\sim 4$ can be achieved for some masses by sliding down the centre of the window. We have also verified that a sliding window analysis of the H.E.S.S.~data with a simple power-law background reproduces our limits in Fig.~\ref{fig:upperlimits} (and thus the published limits in Ref.~\cite{2013PhRvL.110d1301A}) within a factor of $\sim2$, which is remarkable given the more complex background model in the full analysis.

\par Lastly, let us notice that in all searches carried out we have found no significant evidence for the presence of triangular features in the gamma-ray data. In a few occurrences below $10\,$GeV masses, the TS value surpassed 23.7 corresponding to a local significance of $>5\sigma$ (see e.g.~\cite{VertongenWeniger}). Not only is the global significance of these occurrences small given the large number of trials, but also the effect has smeared for the narrower window with $\epsilon=1.5$.

\section{Spin polarization of the intermediate fermion}\label{app:spin}

\begin{figure}[t!]
\center
\hspace{-0.15cm}\includegraphics[width=0.49\textwidth]{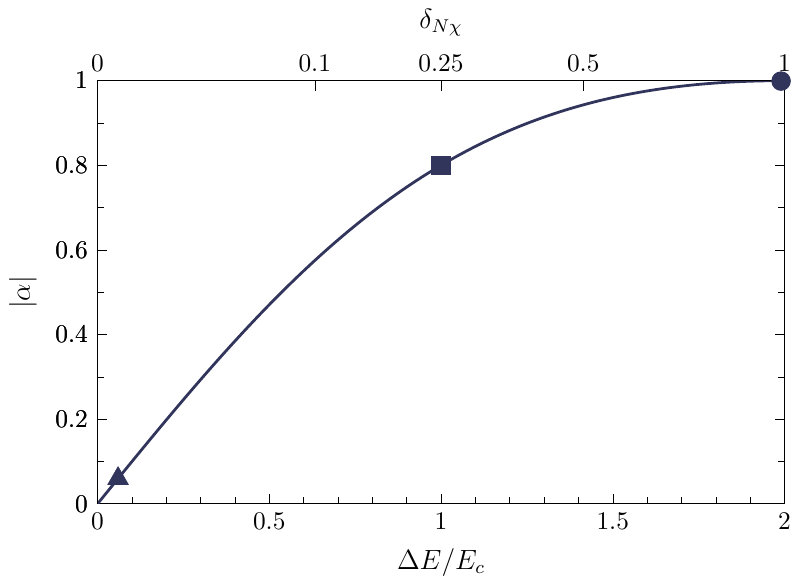}
\caption{\small Correlation between the spin polarization $\alpha$ of the decaying fermion $N_D$ and the relative width of the photon spectrum. We impose gamma-ray constraints on our simplified asymmetric dark matter model for the three benchmarks indicated in the plot: $(\dNc,\alpha)=(0.25,0.8)$ (square), $(\dNc,\alpha)=(0.99,\approx1)$ (circle) and $(\dNc,\alpha)=(0.001,0.06\approx0)$ (triangle). The photon spectra corresponding to these configurations are displayed in Fig.~\ref{fig:spectra1} (right). Notice that for the third benchmark we assume a vanishing $\alpha$ for simplicity; this approximation has little impact on our results.
} \label{fig:alphawidh}
\end{figure}

\par We report  the calculation of the intermediate fermion spin polarization $\alpha$ in the simplified model depicted in Sec.~\ref{sec:model}. This quantity determines the shape of the photon spectrum in Eq.~\eqref{eq:dNdx}. We consider the interaction Lagrangian given in Eq.~\eqref{Lagr1} with the left-handed chiral projector $P_L$. In this case, the annihilation of dark matter particles $\chi$ mostly produces right-handed (positive-helicity) Dirac fermions $N_D$. In fact, in the case of fully polarized fermions in the initial and final states, the annihilation cross sections at leading order in the dark matter velocity are
\begin{eqnarray}
	&& \sigma v_{\times\times,--}=
	\frac{f^4}{64 \pi m_\chi^2} \frac{\left(1- \sqrt{\dNc }\right)^2\sqrt{\dNc}}{\left(1 - \dSc + \dNc \right)^2}\,,\label{cspol1}\\
	&& \sigma v_{\times\times,++}= \frac{f^4}{64 \pi m_\chi^2} \frac{\left(1+ \sqrt{\dNc }\right)^2\sqrt{\dNc}}{\left(1 - \dSc + \dNc \right)^2} \, ,\label{cspol2}
\end{eqnarray}
where the plus (minus) sign refers to positive-helicity (negative-helicity) states of the initial and final particles and the cross represents either helicity.

\par The radiative decay of a polarized Dirac fermion $N_D$ produces a flux of photons with the distribution given in Eq.~\eqref{eq:ftheta} and $\alpha=-1$ ($\alpha=1$) for the decay of a positive-helicity (negative-helicity) state. Then, taking into account Eqs.~\eqref{cspol1} and \eqref{cspol2}, the probability $\mathcal{P}_\pm$ of producing a fermion with positive (negative) helicity is independent of the mass of the scalar mediator and is given by
\begin{equation}\label{prpol}
	\mathcal{P}_\pm  =  \frac 1 4 \frac{\sigma v_{--,\pm \pm}+\sigma v_{++,\pm\pm}  }{( \sigma v)_0}= \frac 1 2  \pm \frac{\sqrt{\dNc}}{1+\dNc} \, ,
\end{equation}
where the unpolarized cross section $( \sigma v)_0$ is reported in Eq.~\eqref{csunpol}. Therefore, the resulting photon spectrum is
\begin{eqnarray}
\frac{\D f}{\D \cos{\theta'}} & = & \frac{1}{2} \mathcal{P}_+\left( 1 - \cos{\theta'}\right) + \frac{1}{2} \mathcal{P}_-\left( 1 +  \cos{\theta'}\right)\nonumber\\
					& \equiv & \frac{1}{2} \left( 1 + \alpha \cos{\theta'}\right) \, ,
\end{eqnarray}
where $\theta'$ is the angle between the photon momentum in the rest frame of $N_D$ and the direction of motion of $N_D$ in the laboratory frame, whereas the spin polarization $\alpha$ reads
\begin{equation}
	\alpha = -\frac{2 \sqrt{\dNc}}{1 + \dNc} \, .\label{alphamodel}
\end{equation}
In the case of a right-handed chiral projector in Eq.~\eqref{Lagr1}, the computation proceeds in the same way and the resulting spin polarization is given by the expression in Eq.~\eqref{alphamodel} with opposite sign.

\par We report in Fig.~\ref{fig:alphawidh} the correlation between the relative width of the photon spectrum, cf.~Eq.~\eqref{eq:centralwidth}, $|\alpha|$ and  $\delta_{N\chi}$. We highlight in the figure three benchmark points corresponding to specific ($\delta_{N\chi}, \alpha$) pairs. These values  are used to set the model-independent constraints on the annihilation cross section $( \sigma v )^{2\gamma}_0$ in the right plot of Fig.~\ref{fig:upperlimits} and the limits on the model parameter space shown in Fig.~\ref{fig:flimits}.

\bibliographystyle{apsrev}
\bibliography{trispec}

\begin{thebibliography}{70}
\expandafter\ifx\csname natexlab\endcsname\relax\def\natexlab#1{#1}\fi
\expandafter\ifx\csname bibnamefont\endcsname\relax
  \def\bibnamefont#1{#1}\fi
\expandafter\ifx\csname bibfnamefont\endcsname\relax
  \def\bibfnamefont#1{#1}\fi
\expandafter\ifx\csname citenamefont\endcsname\relax
  \def\citenamefont#1{#1}\fi
\expandafter\ifx\csname url\endcsname\relax
  \def\url#1{\texttt{#1}}\fi
\expandafter\ifx\csname urlprefix\endcsname\relax\def\urlprefix{URL }\fi
\providecommand{\bibinfo}[2]{#2}
\providecommand{\eprint}[2][]{\url{#2}}

\bibitem[{\citenamefont{Ade et~al.}(2014)}]{Ade:2013zuv}
\bibinfo{author}{\bibfnamefont{P.~A.~R.} \bibnamefont{Ade}}
  \bibnamefont{et~al.} (\bibinfo{collaboration}{Planck}),
  \bibinfo{journal}{Astron. Astrophys.} \textbf{\bibinfo{volume}{571}},
  \bibinfo{pages}{A16} (\bibinfo{year}{2014}), \eprint{1303.5076}.

\bibitem[{\citenamefont{Bertone et~al.}(2005)\citenamefont{Bertone, Hooper, and
  Silk}}]{BertoneReview}
\bibinfo{author}{\bibfnamefont{G.}~\bibnamefont{Bertone}},
  \bibinfo{author}{\bibfnamefont{D.}~\bibnamefont{Hooper}}, \bibnamefont{and}
  \bibinfo{author}{\bibfnamefont{J.}~\bibnamefont{Silk}},
  \bibinfo{journal}{Phys. Rept.} \textbf{\bibinfo{volume}{405}},
  \bibinfo{pages}{279} (\bibinfo{year}{2005}), \eprint{hep-ph/0404175}.

\bibitem[{\citenamefont{Bergstrom}(2000)}]{BergstromReview}
\bibinfo{author}{\bibfnamefont{L.}~\bibnamefont{Bergstrom}},
  \bibinfo{journal}{Rept. Prog. Phys.} \textbf{\bibinfo{volume}{63}},
  \bibinfo{pages}{793} (\bibinfo{year}{2000}), \eprint{hep-ph/0002126}.

\bibitem[{\citenamefont{Feng}(2010)}]{Feng:2010gw}
\bibinfo{author}{\bibfnamefont{J.~L.} \bibnamefont{Feng}},
  \bibinfo{journal}{Ann. Rev. Astron. Astrophys.}
  \textbf{\bibinfo{volume}{48}}, \bibinfo{pages}{495} (\bibinfo{year}{2010}),
  \eprint{1003.0904}.

\bibitem[{\citenamefont{Dine and Kusenko}(2003)}]{Dine:2003ax}
\bibinfo{author}{\bibfnamefont{M.}~\bibnamefont{Dine}} \bibnamefont{and}
  \bibinfo{author}{\bibfnamefont{A.}~\bibnamefont{Kusenko}},
  \bibinfo{journal}{Rev. Mod. Phys.} \textbf{\bibinfo{volume}{76}},
  \bibinfo{pages}{1} (\bibinfo{year}{2003}), \eprint{hep-ph/0303065}.

\bibitem[{\citenamefont{Sakharov}(1967)}]{Sakharov:1967dj}
\bibinfo{author}{\bibfnamefont{A.~D.} \bibnamefont{Sakharov}},
  \bibinfo{journal}{Pisma Zh. Eksp. Teor. Fiz.} \textbf{\bibinfo{volume}{5}},
  \bibinfo{pages}{32} (\bibinfo{year}{1967}), \bibinfo{note}{[Usp. Fiz.
  Nauk161,61(1991)]}.

\bibitem[{\citenamefont{Rubakov and Shaposhnikov}(1996)}]{Rubakov:1996vz}
\bibinfo{author}{\bibfnamefont{V.~A.} \bibnamefont{Rubakov}} \bibnamefont{and}
  \bibinfo{author}{\bibfnamefont{M.~E.} \bibnamefont{Shaposhnikov}},
  \bibinfo{journal}{Usp. Fiz. Nauk} \textbf{\bibinfo{volume}{166}},
  \bibinfo{pages}{493} (\bibinfo{year}{1996}), \bibinfo{note}{[Phys.
  Usp.39,461(1996)]}, \eprint{hep-ph/9603208}.

\bibitem[{\citenamefont{Fukugita and Yanagida}(1986)}]{Fukugita:1986hr}
\bibinfo{author}{\bibfnamefont{M.}~\bibnamefont{Fukugita}} \bibnamefont{and}
  \bibinfo{author}{\bibfnamefont{T.}~\bibnamefont{Yanagida}},
  \bibinfo{journal}{Phys. Lett.} \textbf{\bibinfo{volume}{B174}},
  \bibinfo{pages}{45} (\bibinfo{year}{1986}).

\bibitem[{\citenamefont{Affleck and Dine}(1985)}]{Affleck:1984fy}
\bibinfo{author}{\bibfnamefont{I.}~\bibnamefont{Affleck}} \bibnamefont{and}
  \bibinfo{author}{\bibfnamefont{M.}~\bibnamefont{Dine}},
  \bibinfo{journal}{Nucl. Phys.} \textbf{\bibinfo{volume}{B249}},
  \bibinfo{pages}{361} (\bibinfo{year}{1985}).

\bibitem[{\citenamefont{{Zeldovich}}(1965)}]{Zeldovich}
\bibinfo{author}{\bibfnamefont{Y.~B.} \bibnamefont{{Zeldovich}}},
  \bibinfo{journal}{Adv. Astr. Astrophys.} \textbf{\bibinfo{volume}{3}},
  \bibinfo{pages}{241} (\bibinfo{year}{1965}).

\bibitem[{\citenamefont{Chiu}(1966)}]{Chiu:1966kg}
\bibinfo{author}{\bibfnamefont{H.-Y.} \bibnamefont{Chiu}},
  \bibinfo{journal}{Phys. Rev. Lett.} \textbf{\bibinfo{volume}{17}},
  \bibinfo{pages}{712} (\bibinfo{year}{1966}).

\bibitem[{\citenamefont{Steigman}(1979)}]{Steigman:1979kw}
\bibinfo{author}{\bibfnamefont{G.}~\bibnamefont{Steigman}},
  \bibinfo{journal}{Ann. Rev. Nucl. Part. Sci.} \textbf{\bibinfo{volume}{29}},
  \bibinfo{pages}{313} (\bibinfo{year}{1979}).

\bibitem[{\citenamefont{Scherrer and Turner}(1986)}]{Scherrer:1985zt}
\bibinfo{author}{\bibfnamefont{R.~J.} \bibnamefont{Scherrer}} \bibnamefont{and}
  \bibinfo{author}{\bibfnamefont{M.~S.} \bibnamefont{Turner}},
  \bibinfo{journal}{Phys. Rev.} \textbf{\bibinfo{volume}{D33}},
  \bibinfo{pages}{1585} (\bibinfo{year}{1986}), \bibinfo{note}{[Erratum: Phys.
  Rev.D34,3263(1986)]}.

\bibitem[{\citenamefont{Nussinov}(1985)}]{Nussinov:1985xr}
\bibinfo{author}{\bibfnamefont{S.}~\bibnamefont{Nussinov}},
  \bibinfo{journal}{Phys. Lett.} \textbf{\bibinfo{volume}{B165}},
  \bibinfo{pages}{55} (\bibinfo{year}{1985}).

\bibitem[{\citenamefont{Zurek}(2014)}]{Zurek:2013wia}
\bibinfo{author}{\bibfnamefont{K.~M.} \bibnamefont{Zurek}},
  \bibinfo{journal}{Phys. Rept.} \textbf{\bibinfo{volume}{537}},
  \bibinfo{pages}{91} (\bibinfo{year}{2014}), \eprint{1308.0338}.

\bibitem[{\citenamefont{Petraki and Volkas}(2013)}]{Petraki:2013wwa}
\bibinfo{author}{\bibfnamefont{K.}~\bibnamefont{Petraki}} \bibnamefont{and}
  \bibinfo{author}{\bibfnamefont{R.~R.} \bibnamefont{Volkas}},
  \bibinfo{journal}{Int. J. Mod. Phys.} \textbf{\bibinfo{volume}{A28}},
  \bibinfo{pages}{1330028} (\bibinfo{year}{2013}), \eprint{1305.4939}.

\bibitem[{\citenamefont{Aartsen et~al.}(2015)}]{Aartsen:2015xej}
\bibinfo{author}{\bibfnamefont{M.~G.} \bibnamefont{Aartsen}}
  \bibnamefont{et~al.} (\bibinfo{collaboration}{IceCube}),
  \bibinfo{journal}{Eur. Phys. J.} \textbf{\bibinfo{volume}{C75}},
  \bibinfo{pages}{492} (\bibinfo{year}{2015}), \eprint{1505.07259}.

\bibitem[{\citenamefont{Adrian-Martinez
  et~al.}(2015)}]{Adrian-Martinez:2015wey}
\bibinfo{author}{\bibfnamefont{S.}~\bibnamefont{Adrian-Martinez}}
  \bibnamefont{et~al.} (\bibinfo{collaboration}{ANTARES}),
  \bibinfo{journal}{JCAP} \textbf{\bibinfo{volume}{1510}}, \bibinfo{pages}{068}
  (\bibinfo{year}{2015}), \eprint{1505.04866}.

\bibitem[{\citenamefont{Wendell}(2015)}]{Wendell:2014dka}
\bibinfo{author}{\bibfnamefont{R.}~\bibnamefont{Wendell}}
  (\bibinfo{collaboration}{Super-Kamiokande}), \bibinfo{journal}{AIP Conf.
  Proc.} \textbf{\bibinfo{volume}{1666}}, \bibinfo{pages}{100001}
  (\bibinfo{year}{2015}), \eprint{1412.5234}.

\bibitem[{\citenamefont{Tang}(2015)}]{Tang:2015meg}
\bibinfo{author}{\bibfnamefont{Y.}~\bibnamefont{Tang}} (\bibinfo{year}{2015}),
  \eprint{1512.03159}.

\bibitem[{\citenamefont{Srednicki et~al.}(1986)\citenamefont{Srednicki,
  Theisen, and Silk}}]{Srednicki:1985sf}
\bibinfo{author}{\bibfnamefont{M.}~\bibnamefont{Srednicki}},
  \bibinfo{author}{\bibfnamefont{S.}~\bibnamefont{Theisen}}, \bibnamefont{and}
  \bibinfo{author}{\bibfnamefont{J.}~\bibnamefont{Silk}},
  \bibinfo{journal}{Phys. Rev. Lett.} \textbf{\bibinfo{volume}{56}},
  \bibinfo{pages}{263} (\bibinfo{year}{1986}), \bibinfo{note}{[Erratum: Phys.
  Rev. Lett.56,1883(1986)]}.

\bibitem[{\citenamefont{Rudaz}(1986)}]{Rudaz:1986db}
\bibinfo{author}{\bibfnamefont{S.}~\bibnamefont{Rudaz}},
  \bibinfo{journal}{Phys. Rev. Lett.} \textbf{\bibinfo{volume}{56}},
  \bibinfo{pages}{2128} (\bibinfo{year}{1986}).

\bibitem[{\citenamefont{Bergstrom and Snellman}(1988)}]{Bergstrom:1988fp}
\bibinfo{author}{\bibfnamefont{L.}~\bibnamefont{Bergstrom}} \bibnamefont{and}
  \bibinfo{author}{\bibfnamefont{H.}~\bibnamefont{Snellman}},
  \bibinfo{journal}{Phys. Rev.} \textbf{\bibinfo{volume}{D37}},
  \bibinfo{pages}{3737} (\bibinfo{year}{1988}).

\bibitem[{\citenamefont{Bergstrom}(1989)}]{Bergstrom:1989jr}
\bibinfo{author}{\bibfnamefont{L.}~\bibnamefont{Bergstrom}},
  \bibinfo{journal}{Phys. Lett.} \textbf{\bibinfo{volume}{B225}},
  \bibinfo{pages}{372} (\bibinfo{year}{1989}).

\bibitem[{\citenamefont{Flores et~al.}(1989)\citenamefont{Flores, Olive, and
  Rudaz}}]{Flores:1989ru}
\bibinfo{author}{\bibfnamefont{R.}~\bibnamefont{Flores}},
  \bibinfo{author}{\bibfnamefont{K.~A.} \bibnamefont{Olive}}, \bibnamefont{and}
  \bibinfo{author}{\bibfnamefont{S.}~\bibnamefont{Rudaz}},
  \bibinfo{journal}{Phys. Lett.} \textbf{\bibinfo{volume}{B232}},
  \bibinfo{pages}{377} (\bibinfo{year}{1989}).

\bibitem[{\citenamefont{Bringmann et~al.}(2008)\citenamefont{Bringmann,
  Bergstrom, and Edsjo}}]{Bringmann:2007nk}
\bibinfo{author}{\bibfnamefont{T.}~\bibnamefont{Bringmann}},
  \bibinfo{author}{\bibfnamefont{L.}~\bibnamefont{Bergstrom}},
  \bibnamefont{and} \bibinfo{author}{\bibfnamefont{J.}~\bibnamefont{Edsjo}},
  \bibinfo{journal}{JHEP} \textbf{\bibinfo{volume}{01}}, \bibinfo{pages}{049}
  (\bibinfo{year}{2008}), \eprint{0710.3169}.

\bibitem[{\citenamefont{Ibarra et~al.}(2012)\citenamefont{Ibarra, Lopez~Gehler,
  and Pato}}]{Box2012}
\bibinfo{author}{\bibfnamefont{A.}~\bibnamefont{Ibarra}},
  \bibinfo{author}{\bibfnamefont{S.}~\bibnamefont{Lopez~Gehler}},
  \bibnamefont{and} \bibinfo{author}{\bibfnamefont{M.}~\bibnamefont{Pato}},
  \bibinfo{journal}{JCAP} \textbf{\bibinfo{volume}{1207}}, \bibinfo{pages}{043}
  (\bibinfo{year}{2012}), \eprint{1205.0007}.

\bibitem[{\citenamefont{{Raffelt}}(1996)}]{Raffelt:1996wa}
\bibinfo{author}{\bibfnamefont{G.}~\bibnamefont{{Raffelt}}},
  \emph{\bibinfo{title}{{Stars as laboratories for fundamental physics}}}
  (\bibinfo{publisher}{The University of Chicago Press}, \bibinfo{year}{1996}).

\bibitem[{\citenamefont{{Ibarra} et~al.}()\citenamefont{{Ibarra},
  {L{\'o}pez-Gehler}, {Molinaro}, and {Pato}}}]{NextPaper}
\bibinfo{author}{\bibfnamefont{A.}~\bibnamefont{{Ibarra}}},
  \bibinfo{author}{\bibfnamefont{S.}~\bibnamefont{{L{\'o}pez-Gehler}}},
  \bibinfo{author}{\bibfnamefont{E.}~\bibnamefont{{Molinaro}}},
  \bibnamefont{and} \bibinfo{author}{\bibfnamefont{M.}~\bibnamefont{{Pato}}},
  \bibinfo{note}{{in progress}}.

\bibitem[{\citenamefont{Ibarra et~al.}(2013)\citenamefont{Ibarra, Lee,
  López~Gehler, Park, and Pato}}]{Box2013}
\bibinfo{author}{\bibfnamefont{A.}~\bibnamefont{Ibarra}},
  \bibinfo{author}{\bibfnamefont{H.~M.} \bibnamefont{Lee}},
  \bibinfo{author}{\bibfnamefont{S.}~\bibnamefont{López~Gehler}},
  \bibinfo{author}{\bibfnamefont{W.-I.} \bibnamefont{Park}}, \bibnamefont{and}
  \bibinfo{author}{\bibfnamefont{M.}~\bibnamefont{Pato}},
  \bibinfo{journal}{JCAP} \textbf{\bibinfo{volume}{1305}}, \bibinfo{pages}{016}
  (\bibinfo{year}{2013}), \eprint{1303.6632}.

\bibitem[{\citenamefont{Navarro et~al.}(2004)\citenamefont{Navarro, Hayashi,
  Power, Jenkins, Frenk et~al.}}]{Navarro:2003ew}
\bibinfo{author}{\bibfnamefont{J.~F.} \bibnamefont{Navarro}},
  \bibinfo{author}{\bibfnamefont{E.}~\bibnamefont{Hayashi}},
  \bibinfo{author}{\bibfnamefont{C.}~\bibnamefont{Power}},
  \bibinfo{author}{\bibfnamefont{A.}~\bibnamefont{Jenkins}},
  \bibinfo{author}{\bibfnamefont{C.~S.} \bibnamefont{Frenk}},
  \bibnamefont{et~al.}, \bibinfo{journal}{Mon.Not.Roy.Astron.Soc.}
  \textbf{\bibinfo{volume}{349}}, \bibinfo{pages}{1039} (\bibinfo{year}{2004}),
  \eprint{astro-ph/0311231}.

\bibitem[{\citenamefont{{Merritt} et~al.}(2006)\citenamefont{{Merritt},
  {Graham}, {Moore}, {Diemand}, and {Terzi{\'c}}}}]{Merritt2006}
\bibinfo{author}{\bibfnamefont{D.}~\bibnamefont{{Merritt}}},
  \bibinfo{author}{\bibfnamefont{A.~W.} \bibnamefont{{Graham}}},
  \bibinfo{author}{\bibfnamefont{B.}~\bibnamefont{{Moore}}},
  \bibinfo{author}{\bibfnamefont{J.}~\bibnamefont{{Diemand}}},
  \bibnamefont{and}
  \bibinfo{author}{\bibfnamefont{B.}~\bibnamefont{{Terzi{\'c}}}},
  \bibinfo{journal}{\aj} \textbf{\bibinfo{volume}{132}}, \bibinfo{pages}{2685}
  (\bibinfo{year}{2006}), \eprint{astro-ph/0509417}.

\bibitem[{\citenamefont{{Catena} and {Ullio}}(2010)}]{CatenaUllio2010}
\bibinfo{author}{\bibfnamefont{R.}~\bibnamefont{{Catena}}} \bibnamefont{and}
  \bibinfo{author}{\bibfnamefont{P.}~\bibnamefont{{Ullio}}},
  \bibinfo{journal}{\jcap} \textbf{\bibinfo{volume}{8}}, \bibinfo{eid}{004}
  (\bibinfo{year}{2010}), \eprint{0907.0018}.

\bibitem[{\citenamefont{Weber and de~Boer}(2010)}]{Weber:2009pt}
\bibinfo{author}{\bibfnamefont{M.}~\bibnamefont{Weber}} \bibnamefont{and}
  \bibinfo{author}{\bibfnamefont{W.}~\bibnamefont{de~Boer}},
  \bibinfo{journal}{Astron.Astrophys.} \textbf{\bibinfo{volume}{509}},
  \bibinfo{pages}{A25} (\bibinfo{year}{2010}), \eprint{0910.4272}.

\bibitem[{\citenamefont{Salucci et~al.}(2010)\citenamefont{Salucci, Nesti,
  Gentile, and Martins}}]{Salucci:2010qr}
\bibinfo{author}{\bibfnamefont{P.}~\bibnamefont{Salucci}},
  \bibinfo{author}{\bibfnamefont{F.}~\bibnamefont{Nesti}},
  \bibinfo{author}{\bibfnamefont{G.}~\bibnamefont{Gentile}}, \bibnamefont{and}
  \bibinfo{author}{\bibfnamefont{C.}~\bibnamefont{Martins}},
  \bibinfo{journal}{Astron.Astrophys.} \textbf{\bibinfo{volume}{523}},
  \bibinfo{pages}{A83} (\bibinfo{year}{2010}), \eprint{1003.3101}.

\bibitem[{\citenamefont{{McMillan}}(2011)}]{2011MNRAS.414.2446M}
\bibinfo{author}{\bibfnamefont{P.~J.} \bibnamefont{{McMillan}}},
  \bibinfo{journal}{\mnras} \textbf{\bibinfo{volume}{414}},
  \bibinfo{pages}{2446} (\bibinfo{year}{2011}), \eprint{1102.4340}.

\bibitem[{\citenamefont{{Iocco} et~al.}(2011)\citenamefont{{Iocco}, {Pato},
  {Bertone}, and {Jetzer}}}]{Iocco2011}
\bibinfo{author}{\bibfnamefont{F.}~\bibnamefont{{Iocco}}},
  \bibinfo{author}{\bibfnamefont{M.}~\bibnamefont{{Pato}}},
  \bibinfo{author}{\bibfnamefont{G.}~\bibnamefont{{Bertone}}},
  \bibnamefont{and} \bibinfo{author}{\bibfnamefont{P.}~\bibnamefont{{Jetzer}}},
  \bibinfo{journal}{\jcap} \textbf{\bibinfo{volume}{11}}, \bibinfo{eid}{029}
  (\bibinfo{year}{2011}), \eprint{1107.5810}.

\bibitem[{\citenamefont{Nesti and Salucci}(2013)}]{Nesti:2013uwa}
\bibinfo{author}{\bibfnamefont{F.}~\bibnamefont{Nesti}} \bibnamefont{and}
  \bibinfo{author}{\bibfnamefont{P.}~\bibnamefont{Salucci}},
  \bibinfo{journal}{JCAP} \textbf{\bibinfo{volume}{1307}}, \bibinfo{pages}{016}
  (\bibinfo{year}{2013}), \eprint{1304.5127}.

\bibitem[{\citenamefont{{Gillessen} et~al.}(2009)\citenamefont{{Gillessen},
  {Eisenhauer}, {Trippe}, {Alexander}, {Genzel}, {Martins}, and
  {Ott}}}]{Gillessen2009}
\bibinfo{author}{\bibfnamefont{S.}~\bibnamefont{{Gillessen}}},
  \bibinfo{author}{\bibfnamefont{F.}~\bibnamefont{{Eisenhauer}}},
  \bibinfo{author}{\bibfnamefont{S.}~\bibnamefont{{Trippe}}},
  \bibinfo{author}{\bibfnamefont{T.}~\bibnamefont{{Alexander}}},
  \bibinfo{author}{\bibfnamefont{R.}~\bibnamefont{{Genzel}}},
  \bibinfo{author}{\bibfnamefont{F.}~\bibnamefont{{Martins}}},
  \bibnamefont{and} \bibinfo{author}{\bibfnamefont{T.}~\bibnamefont{{Ott}}},
  \bibinfo{journal}{\apj} \textbf{\bibinfo{volume}{692}}, \bibinfo{pages}{1075}
  (\bibinfo{year}{2009}), \eprint{0810.4674}.

\bibitem[{\citenamefont{{Ando} et~al.}(2011)\citenamefont{{Ando}, {Nagayama},
  {Omodaka}, {Handa}, {Imai}, {Nakagawa}, {Nakanishi}, {Honma}, {Kobayashi},
  and {Miyaji}}}]{Ando2011}
\bibinfo{author}{\bibfnamefont{K.}~\bibnamefont{{Ando}}},
  \bibinfo{author}{\bibfnamefont{T.}~\bibnamefont{{Nagayama}}},
  \bibinfo{author}{\bibfnamefont{T.}~\bibnamefont{{Omodaka}}},
  \bibinfo{author}{\bibfnamefont{T.}~\bibnamefont{{Handa}}},
  \bibinfo{author}{\bibfnamefont{H.}~\bibnamefont{{Imai}}},
  \bibinfo{author}{\bibfnamefont{A.}~\bibnamefont{{Nakagawa}}},
  \bibinfo{author}{\bibfnamefont{H.}~\bibnamefont{{Nakanishi}}},
  \bibinfo{author}{\bibfnamefont{M.}~\bibnamefont{{Honma}}},
  \bibinfo{author}{\bibfnamefont{H.}~\bibnamefont{{Kobayashi}}},
  \bibnamefont{and} \bibinfo{author}{\bibfnamefont{T.}~\bibnamefont{{Miyaji}}},
  \bibinfo{journal}{\pasj} \textbf{\bibinfo{volume}{63}}, \bibinfo{pages}{45}
  (\bibinfo{year}{2011}), \eprint{1012.5715}.

\bibitem[{\citenamefont{{Malkin}}(2012)}]{Malkin2012}
\bibinfo{author}{\bibfnamefont{Z.}~\bibnamefont{{Malkin}}},
  \bibinfo{journal}{ArXiv e-prints}  (\bibinfo{year}{2012}),
  \eprint{1202.6128}.

\bibitem[{\citenamefont{{Reid} et~al.}(2014)\citenamefont{{Reid}, {Menten},
  {Brunthaler}, {Zheng}, {Dame}, {Xu}, {Wu}, {Zhang}, {Sanna}, {Sato}
  et~al.}}]{Reid2014}
\bibinfo{author}{\bibfnamefont{M.~J.} \bibnamefont{{Reid}}},
  \bibinfo{author}{\bibfnamefont{K.~M.} \bibnamefont{{Menten}}},
  \bibinfo{author}{\bibfnamefont{A.}~\bibnamefont{{Brunthaler}}},
  \bibinfo{author}{\bibfnamefont{X.~W.} \bibnamefont{{Zheng}}},
  \bibinfo{author}{\bibfnamefont{T.~M.} \bibnamefont{{Dame}}},
  \bibinfo{author}{\bibfnamefont{Y.}~\bibnamefont{{Xu}}},
  \bibinfo{author}{\bibfnamefont{Y.}~\bibnamefont{{Wu}}},
  \bibinfo{author}{\bibfnamefont{B.}~\bibnamefont{{Zhang}}},
  \bibinfo{author}{\bibfnamefont{A.}~\bibnamefont{{Sanna}}},
  \bibinfo{author}{\bibfnamefont{M.}~\bibnamefont{{Sato}}},
  \bibnamefont{et~al.}, \bibinfo{journal}{\apj} \textbf{\bibinfo{volume}{783}},
  \bibinfo{eid}{130} (\bibinfo{year}{2014}), \eprint{1401.5377}.

\bibitem[{\citenamefont{{Ibarra} et~al.}(2015)\citenamefont{{Ibarra},
  {Lamperstorfer}, {L{\'o}pez-Gehler}, {Pato}, and
  {Bertone}}}]{2015JCAP...09..048I}
\bibinfo{author}{\bibfnamefont{A.}~\bibnamefont{{Ibarra}}},
  \bibinfo{author}{\bibfnamefont{A.~S.} \bibnamefont{{Lamperstorfer}}},
  \bibinfo{author}{\bibfnamefont{S.}~\bibnamefont{{L{\'o}pez-Gehler}}},
  \bibinfo{author}{\bibfnamefont{M.}~\bibnamefont{{Pato}}}, \bibnamefont{and}
  \bibinfo{author}{\bibfnamefont{G.}~\bibnamefont{{Bertone}}},
  \bibinfo{journal}{\jcap} \textbf{\bibinfo{volume}{9}}, \bibinfo{eid}{048}
  (\bibinfo{year}{2015}), \eprint{1503.06797}.

\bibitem[{\citenamefont{Iminniyaz et~al.}(2011)\citenamefont{Iminniyaz, Drees,
  and Chen}}]{Iminniyaz:2011yp}
\bibinfo{author}{\bibfnamefont{H.}~\bibnamefont{Iminniyaz}},
  \bibinfo{author}{\bibfnamefont{M.}~\bibnamefont{Drees}}, \bibnamefont{and}
  \bibinfo{author}{\bibfnamefont{X.}~\bibnamefont{Chen}},
  \bibinfo{journal}{JCAP} \textbf{\bibinfo{volume}{1107}}, \bibinfo{pages}{003}
  (\bibinfo{year}{2011}), \eprint{1104.5548}.

\bibitem[{\citenamefont{Falkowski et~al.}(2011)\citenamefont{Falkowski,
  Ruderman, and Volansky}}]{Falkowski:2011xh}
\bibinfo{author}{\bibfnamefont{A.}~\bibnamefont{Falkowski}},
  \bibinfo{author}{\bibfnamefont{J.~T.} \bibnamefont{Ruderman}},
  \bibnamefont{and} \bibinfo{author}{\bibfnamefont{T.}~\bibnamefont{Volansky}},
  \bibinfo{journal}{JHEP} \textbf{\bibinfo{volume}{05}}, \bibinfo{pages}{106}
  (\bibinfo{year}{2011}), \eprint{1101.4936}.

\bibitem[{\citenamefont{Kusenko}(2009)}]{Kusenko:2009up}
\bibinfo{author}{\bibfnamefont{A.}~\bibnamefont{Kusenko}},
  \bibinfo{journal}{Phys. Rept.} \textbf{\bibinfo{volume}{481}},
  \bibinfo{pages}{1} (\bibinfo{year}{2009}), \eprint{0906.2968}.

\bibitem[{\citenamefont{Atre et~al.}(2009)\citenamefont{Atre, Han, Pascoli, and
  Zhang}}]{Atre:2009rg}
\bibinfo{author}{\bibfnamefont{A.}~\bibnamefont{Atre}},
  \bibinfo{author}{\bibfnamefont{T.}~\bibnamefont{Han}},
  \bibinfo{author}{\bibfnamefont{S.}~\bibnamefont{Pascoli}}, \bibnamefont{and}
  \bibinfo{author}{\bibfnamefont{B.}~\bibnamefont{Zhang}},
  \bibinfo{journal}{JHEP} \textbf{\bibinfo{volume}{05}}, \bibinfo{pages}{030}
  (\bibinfo{year}{2009}), \eprint{0901.3589}.

\bibitem[{\citenamefont{Deppisch et~al.}(2015)\citenamefont{Deppisch,
  Bhupal~Dev, and Pilaftsis}}]{Deppisch:2015qwa}
\bibinfo{author}{\bibfnamefont{F.~F.} \bibnamefont{Deppisch}},
  \bibinfo{author}{\bibfnamefont{P.~S.} \bibnamefont{Bhupal~Dev}},
  \bibnamefont{and}
  \bibinfo{author}{\bibfnamefont{A.}~\bibnamefont{Pilaftsis}},
  \bibinfo{journal}{New J. Phys.} \textbf{\bibinfo{volume}{17}},
  \bibinfo{pages}{075019} (\bibinfo{year}{2015}), \eprint{1502.06541}.

\bibitem[{\citenamefont{Josse-Michaux and
  Molinaro}(2011)}]{JosseMichaux:2011ba}
\bibinfo{author}{\bibfnamefont{F.-X.} \bibnamefont{Josse-Michaux}}
  \bibnamefont{and} \bibinfo{author}{\bibfnamefont{E.}~\bibnamefont{Molinaro}},
  \bibinfo{journal}{Phys. Rev.} \textbf{\bibinfo{volume}{D84}},
  \bibinfo{pages}{125021} (\bibinfo{year}{2011}), \eprint{1108.0482}.

\bibitem[{\citenamefont{Josse-Michaux and
  Molinaro}(2013)}]{JosseMichaux:2012wj}
\bibinfo{author}{\bibfnamefont{F.-X.} \bibnamefont{Josse-Michaux}}
  \bibnamefont{and} \bibinfo{author}{\bibfnamefont{E.}~\bibnamefont{Molinaro}},
  \bibinfo{journal}{Phys. Rev.} \textbf{\bibinfo{volume}{D87}},
  \bibinfo{pages}{036007} (\bibinfo{year}{2013}), \eprint{1210.7202}.

\bibitem[{\citenamefont{del Aguila and
  Aguilar-Saavedra}(2009)}]{delAguila:2008hw}
\bibinfo{author}{\bibfnamefont{F.}~\bibnamefont{del Aguila}} \bibnamefont{and}
  \bibinfo{author}{\bibfnamefont{J.~A.} \bibnamefont{Aguilar-Saavedra}},
  \bibinfo{journal}{Phys. Lett.} \textbf{\bibinfo{volume}{B672}},
  \bibinfo{pages}{158} (\bibinfo{year}{2009}), \eprint{0809.2096}.

\bibitem[{\citenamefont{Gavela et~al.}(2009)\citenamefont{Gavela, Hambye,
  Hernandez, and Hernandez}}]{Gavela:2009cd}
\bibinfo{author}{\bibfnamefont{M.~B.} \bibnamefont{Gavela}},
  \bibinfo{author}{\bibfnamefont{T.}~\bibnamefont{Hambye}},
  \bibinfo{author}{\bibfnamefont{D.}~\bibnamefont{Hernandez}},
  \bibnamefont{and}
  \bibinfo{author}{\bibfnamefont{P.}~\bibnamefont{Hernandez}},
  \bibinfo{journal}{JHEP} \textbf{\bibinfo{volume}{09}}, \bibinfo{pages}{038}
  (\bibinfo{year}{2009}), \eprint{0906.1461}.

\bibitem[{\citenamefont{Ibarra et~al.}(2010)\citenamefont{Ibarra, Molinaro, and
  Petcov}}]{Ibarra:2010xw}
\bibinfo{author}{\bibfnamefont{A.}~\bibnamefont{Ibarra}},
  \bibinfo{author}{\bibfnamefont{E.}~\bibnamefont{Molinaro}}, \bibnamefont{and}
  \bibinfo{author}{\bibfnamefont{S.~T.} \bibnamefont{Petcov}},
  \bibinfo{journal}{JHEP} \textbf{\bibinfo{volume}{09}}, \bibinfo{pages}{108}
  (\bibinfo{year}{2010}), \eprint{1007.2378}.

\bibitem[{\citenamefont{Ibarra et~al.}(2011)\citenamefont{Ibarra, Molinaro, and
  Petcov}}]{Ibarra:2011xn}
\bibinfo{author}{\bibfnamefont{A.}~\bibnamefont{Ibarra}},
  \bibinfo{author}{\bibfnamefont{E.}~\bibnamefont{Molinaro}}, \bibnamefont{and}
  \bibinfo{author}{\bibfnamefont{S.~T.} \bibnamefont{Petcov}},
  \bibinfo{journal}{Phys. Rev.} \textbf{\bibinfo{volume}{D84}},
  \bibinfo{pages}{013005} (\bibinfo{year}{2011}), \eprint{1103.6217}.

\bibitem[{\citenamefont{Dinh et~al.}(2012)\citenamefont{Dinh, Ibarra, Molinaro,
  and Petcov}}]{Dinh:2012bp}
\bibinfo{author}{\bibfnamefont{D.~N.} \bibnamefont{Dinh}},
  \bibinfo{author}{\bibfnamefont{A.}~\bibnamefont{Ibarra}},
  \bibinfo{author}{\bibfnamefont{E.}~\bibnamefont{Molinaro}}, \bibnamefont{and}
  \bibinfo{author}{\bibfnamefont{S.~T.} \bibnamefont{Petcov}},
  \bibinfo{journal}{JHEP} \textbf{\bibinfo{volume}{08}}, \bibinfo{pages}{125}
  (\bibinfo{year}{2012}), \bibinfo{note}{[Erratum: JHEP09,023(2013)]},
  \eprint{1205.4671}.

\bibitem[{\citenamefont{Lopez-Pavon et~al.}(2015)\citenamefont{Lopez-Pavon,
  Molinaro, and Petcov}}]{Lopez-Pavon:2015cga}
\bibinfo{author}{\bibfnamefont{J.}~\bibnamefont{Lopez-Pavon}},
  \bibinfo{author}{\bibfnamefont{E.}~\bibnamefont{Molinaro}}, \bibnamefont{and}
  \bibinfo{author}{\bibfnamefont{S.~T.} \bibnamefont{Petcov}},
  \bibinfo{journal}{JHEP} \textbf{\bibinfo{volume}{11}}, \bibinfo{pages}{030}
  (\bibinfo{year}{2015}), \eprint{1506.05296}.

\bibitem[{\citenamefont{Pal and Wolfenstein}(1982)}]{Pal:1981rm}
\bibinfo{author}{\bibfnamefont{P.~B.} \bibnamefont{Pal}} \bibnamefont{and}
  \bibinfo{author}{\bibfnamefont{L.}~\bibnamefont{Wolfenstein}},
  \bibinfo{journal}{Phys. Rev.} \textbf{\bibinfo{volume}{D25}},
  \bibinfo{pages}{766} (\bibinfo{year}{1982}).

\bibitem[{\citenamefont{Sjostrand et~al.}(2006)\citenamefont{Sjostrand, Mrenna,
  and Skands}}]{Sjostrand:2006za}
\bibinfo{author}{\bibfnamefont{T.}~\bibnamefont{Sjostrand}},
  \bibinfo{author}{\bibfnamefont{S.}~\bibnamefont{Mrenna}}, \bibnamefont{and}
  \bibinfo{author}{\bibfnamefont{P.~Z.} \bibnamefont{Skands}},
  \bibinfo{journal}{JHEP} \textbf{\bibinfo{volume}{05}}, \bibinfo{pages}{026}
  (\bibinfo{year}{2006}), \eprint{hep-ph/0603175}.

\bibitem[{fer({\natexlab{a}})}]{fermilatsite}
\bibinfo{howpublished}{\url{http://www-glast.stanford.edu/}}.

\bibitem[{\citenamefont{Vertongen and Weniger}(2011)}]{VertongenWeniger}
\bibinfo{author}{\bibfnamefont{G.}~\bibnamefont{Vertongen}} \bibnamefont{and}
  \bibinfo{author}{\bibfnamefont{C.}~\bibnamefont{Weniger}},
  \bibinfo{journal}{JCAP} \textbf{\bibinfo{volume}{1105}}, \bibinfo{pages}{027}
  (\bibinfo{year}{2011}), \eprint{1101.2610}.

\bibitem[{fer({\natexlab{b}})}]{fermilatsite2}
\bibinfo{howpublished}{\url{http://www.slac.stanford.edu/exp/glast/groups/canda/lat_Performance.htm}}.

\bibitem[{fer({\natexlab{c}})}]{fermilatdata1}
\bibinfo{howpublished}{\url{http://fermi.gsfc.nasa.gov/ssc/data/access/}}.

\bibitem[{fer({\natexlab{d}})}]{fermilatdata2}
\bibinfo{howpublished}{\url{http://fermi.gsfc.nasa.gov/ssc/data/analysis/software/}}.

\bibitem[{\citenamefont{Ackermann et~al.}(2013)}]{Ackermann:2013uma}
\bibinfo{author}{\bibfnamefont{M.}~\bibnamefont{Ackermann}}
  \bibnamefont{et~al.} (\bibinfo{collaboration}{Fermi-LAT}),
  \bibinfo{journal}{Phys. Rev.} \textbf{\bibinfo{volume}{D88}},
  \bibinfo{pages}{082002} (\bibinfo{year}{2013}), \eprint{1305.5597}.

\bibitem[{\citenamefont{{Abramowski} et~al.}(2013)\citenamefont{{Abramowski},
  {Acero}, {Aharonian}, {Akhperjanian}, {Anton}, {Balenderan}, {Balzer},
  {Barnacka}, {Becherini}, {Becker Tjus} et~al.}}]{2013PhRvL.110d1301A}
\bibinfo{author}{\bibfnamefont{A.}~\bibnamefont{{Abramowski}}},
  \bibinfo{author}{\bibfnamefont{F.}~\bibnamefont{{Acero}}},
  \bibinfo{author}{\bibfnamefont{F.}~\bibnamefont{{Aharonian}}},
  \bibinfo{author}{\bibfnamefont{A.~G.} \bibnamefont{{Akhperjanian}}},
  \bibinfo{author}{\bibfnamefont{G.}~\bibnamefont{{Anton}}},
  \bibinfo{author}{\bibfnamefont{S.}~\bibnamefont{{Balenderan}}},
  \bibinfo{author}{\bibfnamefont{A.}~\bibnamefont{{Balzer}}},
  \bibinfo{author}{\bibfnamefont{A.}~\bibnamefont{{Barnacka}}},
  \bibinfo{author}{\bibfnamefont{Y.}~\bibnamefont{{Becherini}}},
  \bibinfo{author}{\bibfnamefont{J.}~\bibnamefont{{Becker Tjus}}},
  \bibnamefont{et~al.}, \bibinfo{journal}{Physical Review Letters}
  \textbf{\bibinfo{volume}{110}}, \bibinfo{eid}{041301} (\bibinfo{year}{2013}),
  \eprint{1301.1173}.

\bibitem[{\citenamefont{Rolke et~al.}(2005)\citenamefont{Rolke, Lopez, and
  Conrad}}]{Rolke:2004mj}
\bibinfo{author}{\bibfnamefont{W.~A.} \bibnamefont{Rolke}},
  \bibinfo{author}{\bibfnamefont{A.~M.} \bibnamefont{Lopez}}, \bibnamefont{and}
  \bibinfo{author}{\bibfnamefont{J.}~\bibnamefont{Conrad}},
  \bibinfo{journal}{Nucl.Instrum.Meth.} \textbf{\bibinfo{volume}{A551}},
  \bibinfo{pages}{493} (\bibinfo{year}{2005}), \eprint{physics/0403059}.

\bibitem[{\citenamefont{Bringmann et~al.}(2011)\citenamefont{Bringmann, Calore,
  Vertongen, and Weniger}}]{Bringmann:2011ye}
\bibinfo{author}{\bibfnamefont{T.}~\bibnamefont{Bringmann}},
  \bibinfo{author}{\bibfnamefont{F.}~\bibnamefont{Calore}},
  \bibinfo{author}{\bibfnamefont{G.}~\bibnamefont{Vertongen}},
  \bibnamefont{and} \bibinfo{author}{\bibfnamefont{C.}~\bibnamefont{Weniger}},
  \bibinfo{journal}{Phys.Rev.} \textbf{\bibinfo{volume}{D84}},
  \bibinfo{pages}{103525} (\bibinfo{year}{2011}), \eprint{1106.1874}.

\bibitem[{\citenamefont{Bringmann et~al.}(2012)\citenamefont{Bringmann, Huang,
  Ibarra, Vogl, and Weniger}}]{Bringmannetal}
\bibinfo{author}{\bibfnamefont{T.}~\bibnamefont{Bringmann}},
  \bibinfo{author}{\bibfnamefont{X.}~\bibnamefont{Huang}},
  \bibinfo{author}{\bibfnamefont{A.}~\bibnamefont{Ibarra}},
  \bibinfo{author}{\bibfnamefont{S.}~\bibnamefont{Vogl}}, \bibnamefont{and}
  \bibinfo{author}{\bibfnamefont{C.}~\bibnamefont{Weniger}}
  (\bibinfo{year}{2012}), \eprint{1203.1312}.

\bibitem[{\citenamefont{Weniger}(2012)}]{Weniger:2012tx}
\bibinfo{author}{\bibfnamefont{C.}~\bibnamefont{Weniger}}
  (\bibinfo{year}{2012}), \eprint{1204.2797}.

\bibitem[{\citenamefont{{Cowan}}(1997)}]{1997sda..book.....C}
\bibinfo{author}{\bibfnamefont{G.}~\bibnamefont{{Cowan}}},
  \emph{\bibinfo{title}{{Statistical data analysis}}}
  (\bibinfo{publisher}{Oxford University Press}, \bibinfo{year}{1997}).

\end{thebibliography}

\end{document}